\documentclass[twocolumn]{aastex62}
\received{---}
\revised{----}
\accepted{--- }          

\usepackage{color}
\usepackage{graphicx}
\usepackage{amssymb}
\usepackage{array}
\usepackage{color}
\usepackage{subfigure}
\usepackage{csquotes}
\usepackage{natbib}
\bibpunct{(}{)}{;}{a}{,}{,}

\newcommand{\degree}{\ensuremath{^\circ}}

\def \sun {$_{\odot}$}

\shorttitle{B-fields in G34.43+0.24}
\shortauthors{Soam et al.}

\begin{document}

\title{Magnetic fields in the infrared dark cloud G34.43+0.24}

\correspondingauthor{Archana Soam}
\email{asoam@usra.edu, archanasoam.bhu@gmail.com}

\author[0000-0002-6386-2906]{Archana Soam}
\affiliation{SOFIA Science Centre, USRA, NASA Ames Research Centre, MS-12, N232, Moffett Field, CA 94035, USA}
\affiliation{Korea Astronomy and Space Science Institute, 776 Daedeokdae-ro, Yuseong-gu, Daejeon 34055, Republic of Korea}

\author{Tie Liu}
\affiliation{Shanghai Astronomical Observatory, Chinese Academy of Sciences, 80 Nandan Road, Shanghai 200030, China}
\affiliation{Korea Astronomy and Space Science Institute, 776 Daedeokdae-ro, Yuseong-gu, Daejeon 34055, Republic of Korea}
\affiliation{East Asian Observatory, 660 N. A’ohoku Place, Hilo, HI 96720, USA}

\author{B-G Andersson}
\affiliation{SOFIA Science Centre, USRA, NASA Ames Research Centre, MS-12, N232, Moffett Field, CA 94035, USA}

\author{Chang Won Lee} 
\affiliation{Korea Astronomy and Space Science Institute, 776 Daedeokdae-ro, Yuseong-gu, Daejeon 34055, Republic of Korea}
\affiliation{University of Science and Technology, Korea (UST), 217 Gajeong-ro, Yuseong-gu, Daejeon 34113, Republic of Korea}

\author{Junhao Liu}
\affiliation{School of Astronomy and Space Science, Nanjing University, 163 Xianlin Avenue, Nanjing 210023, People’s Republic of China}
\affiliation{Key Laboratory of Modern Astronomy and Astrophysics (Nanjing University), Ministry of Education, Nanjing 210023, People’s Republic of China}
\affiliation{Harvard-Smithsonian Center for Astrophysics, 60 Garden Street, Cambridge, MA 02138, USA}

\author{Mika Juvela}
\affiliation{Department of Physics, P.O.Box 64, FI-00014, University of Helsinki, Finland}

\author{Pak Shing Li}
\affiliation{University of California, Berkeley, United States}

\author{Paul F. Goldsmith}
\affiliation{Jet Propulsion Laboratory, National Aeronautics and Space Administration, United States}

\author{Qizhou Zhang}
\affiliation{Harvard-Smithsonian Center for Astrophysics, 60 Garden Street, Cambridge, MA 02138, USA}

\author{Patrick M. Koch}
\affiliation{Academia Sinica Institute of Astronomy and Astrophysics, P.O. Box 23-141, Taipei 10617, Taiwan}

\author{Kee-Tae Kim}
\affiliation{Korea Astronomy and Space Science Institute, 776 Daedeokdae-ro, Yuseong-gu, Daejeon 34055, Republic of Korea}

\author{Keping Qiu}
\affiliation{School of Astronomy and Space Science, Nanjing University, 163 Xianlin Avenue, Nanjing 210023, People’s Republic of China}
\affiliation{Key Laboratory of Modern Astronomy and Astrophysics (Nanjing University), Ministry of Education, Nanjing 210023, People’s Republic of China}

\author{Neal J. Evans II}
\affiliation{Korea Astronomy and Space Science Institute, 776 Daedeokdae-ro, Yuseong-gu, Daejeon 34055, Republic of Korea}
\affiliation{Department of Astronomy, The University of Texas at Austin, 2515 Speedway, Stop C1400, Austin, Texas 78712-1205, USA}
\affiliation{Humanitas College, Global Campus, Kyung Hee University, Yongin-shi 17104, Korea}

\author{Doug Johnstone}
\affiliation{NRC Herzberg Astronomy and Astrophysics, 5071 West Saanich Road, Victoria, BC V9E 2E7, Canada}
\affiliation{Department of Physics and Astronomy, University of Victoria, Victoria, BC V8P 1A1, Canada}

\author{Mark Thompson}
\affiliation{University of Hertfordshire (Centre for Astrophysics Research), United Kingdom}

\author{Derek Ward-Thompson}
\affiliation{Jeremiah Horrocks Institute, University of Central Lancashire, Preston PR1 2HE, UK}

\author{James Di Francesco} 
\affiliation{NRC Herzberg Astronomy and Astrophysics, 5071 West Saanich Road, Victoria, BC V9E 2E7, Canada}
\affiliation{Department of Physics and Astronomy, University of Victoria, Victoria, BC V8P 1A1, Canada}

\author{Ya-Wen Tang}
\affiliation{Academia Sinica Institute of Astronomy and Astrophysics, P.O. Box 23-141, Taipei 10617, Taiwan}

\author{Julien Montillaud}
\affiliation{Institut UTINAM - UMR 6213 - CNRS - Univ Bourgogne Franche Comte, OSU THETA, 41bis avenue de l'Observatoire, 25000 Besan\c{c}on, France}

\author[0000-0003-2011-8172]{Gwanjeong Kim}
\affil{Nobeyama Radio Observatory, National Astronomical Observatory of Japan, National Institutes of Natural Sciences, Nobeyama, Minamimaki, Minamisaku, Nagano 384-1305, Japan}

\author{Steve Mairs}
\affiliation{East Asian Observatory, 660 N. A`oh\={o}k\={u} Place, University Park, Hilo, HI 96720, USA}

\author{Patricio Sanhueza}
\affiliation{National Astronomical Observatory of Japan, National Institutes of Natural Sciences, 2-21-1 Osawa, Mitaka, Tokyo 181-8588, Japan}

\author[0000-0001-9333-5608]{Shinyoung Kim}
\affiliation{Korea Astronomy and Space Science Institute, 776 Daedeokdae-ro, Yuseong-gu, Daejeon 34055, Republic of Korea}
\affiliation{University of Science and Technology, Korea (UST), 217 Gajeong-ro, Yuseong-gu, Daejeon 34113, Republic of Korea}

\author{David Berry}
\affiliation{East Asian Observatory, 660 N. A`oh\={o}k\={u} Place, University Park, Hilo, HI 96720, USA}

\author{Michael S. Gordon}
\affiliation{SOFIA Science Centre, USRA, NASA Ames Research Centre, MS-12, N232, Moffett Field, CA 94035, USA}

\author{Ken'ichi Tatematsu}
\affiliation{National Astronomical Observatory of Japan, National Institutes of Natural Sciences, 2-21-1 Osawa, Mitaka, Tokyo 181-8588, Japan}

\author{Sheng-Yuan Liu}
\affiliation{Academia Sinica Institute of Astronomy and Astrophysics, P.O. Box 23-141, Taipei 10617, Taiwan}

\author{Kate Pattle}
\affiliation{Institute of Astronomy and Department of Physics, National Tsing Hua University, Hsinchu 30013, Taiwan}

\author{David Eden}
\affiliation{Astrophysics Research Institute, Liverpool John Moores University, IC2, Liverpool Science Park, 146 Brownlow Hill, Liverpool L3 5RF, UK}

\author{Peregrine M. McGehee}
\affiliation{College of the Canyons, United States}

\author{Ke Wang}
\affiliation{Kavli Institute for Astronomy and Astrophysics, Peking University, 5 Yiheyuan Road, Haidian District, Beijing 100871, China}

\author{I. Ristorcelli}
\affiliation{Institut pour la Recherche en Astrophysique et Planétologie, France}

\author{Sarah F. Graves}
\affiliation{East Asian Observatory, 660 N. A`oh\={o}k\={u} Place, University Park, Hilo, HI 96720, USA}

\author{Dana Alina}
\affiliation{Nazarbayev University (Department of Physics), Kazakhstan}

\author{Kevin M. Lacaille}
\affiliation{Department of Physics and Astronomy, McMaster University, Hamilton, ON L8S 4M1 Canada}
\affiliation{Department of Physics and Atmospheric Science, Dalhousie University, Halifax B3H 4R2, Canada}

\author{Ludovic Montier}
\affiliation{Institut pour la Recherche en Astrophysique et Planétologie, France}

\author{Geumsook Park}
\affiliation{Korea Astronomy and Space Science Institute, 776 Daedeokdae-ro, Yuseong-gu, Daejeon 34055, Republic of Korea}

\author{Woojin Kwon}
\affiliation{Korea Astronomy and Space Science Institute, 776 Daedeokdae-ro, Yuseong-gu, Daejeon 34055, Republic of Korea}
\affiliation{University of Science and Technology, Korea (UST), 217 Gajeong-ro, Yuseong-gu, Daejeon 34113, Republic of Korea}

\author{Eun Jung Chung}
\affiliation{Korea Astronomy and Space Science Institute, 776 Daedeokdae-ro, Yuseong-gu, Daejeon 34055, Republic of Korea}

\author{Veli-Matti Pelkonen}
\affiliation{Department of Physics, P.O.Box 64, FI-00014, University of Helsinki, Finland}
\affiliation{Institut de Ciències del Cosmos, Universitat de Barcelona, IEEC-UB, Martí Franquès 1, E08028 Barcelona}

\author{Elisabetta R. Micelotta}
\affiliation{Department of Physics, P.O.Box 64, FI-00014, University of Helsinki, Finland}

\author{Mika Saajasto}
\affiliation{Department of Physics, P.O.Box 64, FI-00014, University of Helsinki, Finland}

\author{Gary Fuller}
\affiliation{Jodrell Bank Centre for Astrophysics, School of Physics and Astronomy, University of Manchester, Oxford Road, Manchester, M13 9PL, UK}

\begin{abstract}
We present the B-fields mapped in IRDC G34.43+0.24 using 850\,$\mu$m polarized dust emission observed with the POL-2 instrument at JCMT. We examine the magnetic field geometries and strengths in the northern, central, and southern regions of the filament. The overall field geometry is ordered and aligned closely perpendicular to the filament's main axis, particularly in regions containing the central clumps MM1 and MM2, whereas MM3 in the north has field orientations aligned with its major axis. The overall field orientations are uniform at large (POL-2 at 14$\arcsec$ and SHARP at 10$\arcsec$) to small scales (TADPOL at 2.5$\arcsec$ and SMA at 1.5$\arcsec$) in the MM1 and MM2 regions. SHARP/CSO observations in MM3 at 350\,$\mu$m from Tang et al. show a similar trend as seen in our POL-2 observations. TADPOL observations demonstrate a well-defined field geometry in MM1/MM2 consistent with MHD simulations of accreting filaments. We obtained a plane-of-sky magnetic field strength of 470$\pm$190\,$\mu$G, 100$\pm$40\,$\mu$G, and 60$\pm$34\,$\mu$G in the central, northern and southern regions of G34, respectively, using the updated Davis-Chandrasekhar-Fermi relation. The estimated value of field strength, combined with column density and velocity dispersion values available in the literature, suggests G34 to be marginally critical with criticality parameter $\rm \lambda$ values 0.8$\pm$0.4, 1.1$\pm$0.8, and 0.9$\pm$0.5 in the central, northern, and southern regions, respectively. The turbulent motions in G34 are sub-Alfv\'{e}nic with Alfv\'{e}nic Mach numbers of 0.34$\pm$0.13, 0.53$\pm$0.30, and 0.49$\pm$0.26 in the three regions. The observed aligned B-fields in G34.43+0.24 are consistent with theoretical models suggesting that B-fields play an important role in guiding the contraction of the cloud driven by gravity.

\end{abstract}


\keywords{stars: formation — ISM: kinematics and dynamics — ISM: magnetic fields}



\section{Introduction} \label{sec:intro}

Filamentary structures exist in molecular clouds, with sizes ranging from a few to tens of parsecs \citep{2014prpl.conf...27A, wang16}. Recent magnetohydrodynamic (MHD) simulations \citep{2017MNRAS.465.2254K, 2018MNRAS.473.4220L, 2018MNRAS.480.2939G} probing the formation of large-scale filamentary clouds suggest a complex evolutionary process involving the interaction and fragmentation of dense, velocity-coherent, fibers into chains of cores, resembling observations in nearby clouds \citep[e.g., L1495/B213 and Musca cloud;][]{2013A&A...554A..55H, 2016A&A...591A.104H}. The simulations show that global magnetic fields are expected to be roughly perpendicular to the longer axes of dense filamentary clouds. Several velocity coherent fibers are identified inside the clouds and appear to be supportable along the main filament. In 3D MHD simulations of cluster-forming turbulent molecular cloud clumps, \citet{2017MNRAS.465.2254K} found that B-fields are oriented parallel to sub-virial clouds and perpendicular to denser gravitationally-bound clouds.

Recent ideal MHD simulations by \citet{2019MNRAS.485.4509L} found that the magnetic field helps in shaping the long filamentary structures with field orientation perpendicular to the long axis of the filaments. Their simulation produces fibre-like substructures similar to those observed in L1495 \citep{2013A&A...554A..55H}. There are some other MHD simulations available which include magnetic fields in filaments. \citet{2016ApJ...832..143F} presented MHD simulations studying the effect of magnetic fields, gravity and turbulence on the formation of filaments finding that filament width does not depend on the orientation of filament with respect to the magnetic fields in G0.253+0.016 region. A statistical analysis of nearby clouds such as Taurus, Musca, Ophiuchus, and Chameleon has revealed that B-field lines tend to become parallel to the filament long axes at low densities (or \enquote{diffuse} with a few $\rm cm^{-3}$) \citep[e.g.,][] {2011ApJ...741...21C, 2016A&A...590A.110C, 2015A&A...576A.104P, 2016A&A...586A.138P} but are perpendicular to the denser filamentary structures. \citet{2014ApJ...797...99K} presented a statistical analysis of 50 sources (from 4000 independent measurements observed with the SMA and the CSO) on the scales of 0.1 pc to 0.01 pc with densities $\rm \gtrsim 10^{5}~cm^{-3}$. Their analysis of B-fields and intensity gradients reveals that the field orientation is perpendicular to the sources' major axes.


Polarized thermal dust emission at submillimeter wavelengths probes the magnetic field structure in high-density regions. The Radiative Torque Alignment (RAT) theory of grain alignment is currently one of the most promising models to explain the polarization of light towards clouds and cores \citep{1976Ap&SS..43..291D, 1997ApJ...490..273L}. This model predicts the asymmetric non-spherical dust grains rotate due to radiative torque and align with their long axes perpendicular to ambient magnetic field. Due to low angular resolution (e.g., $\sim 5{\arcmin}$ with Planck) or high dust extinction (optical or near-infrared polarimetery), previous studies of magnetic fields in filamentary clouds have been mostly limited to nearby clouds. So far, magnetic fields have only been investigated in a few infrared dark clouds \citep{2015ApJ...799...74P, 2018ApJ...859..151L, 2018A&A...620A..26J, 2018ApJ...869L...5L}. Additional observations with higher angular resolution towards filamentary clouds and cores are still needed.


To this end, we are conducting a series of dust polarization observations toward the brightest filaments identified in the JCMT legacy survey of $\sim$1000 Planck Galactic Cold Clumps (PGCCs), called SCOPE \citep[SCUBA-2 Continuum Observations of Pre-protostellar Evolution;][]{2018ApJS..234...28L, 2019MNRAS.485.2895E}, with the POL-2 polarimeter at the JCMT. The observational results of two PGCCs, G35.49-0.31 (hereafter G35) and G9.62+0.19 (hereafter G9) are published in \citet{2018ApJ...859..151L} and \citet{2018ApJ...869L...5L}, respectively.

\begin{figure}
\resizebox{9.5cm}{12.0cm}{\includegraphics{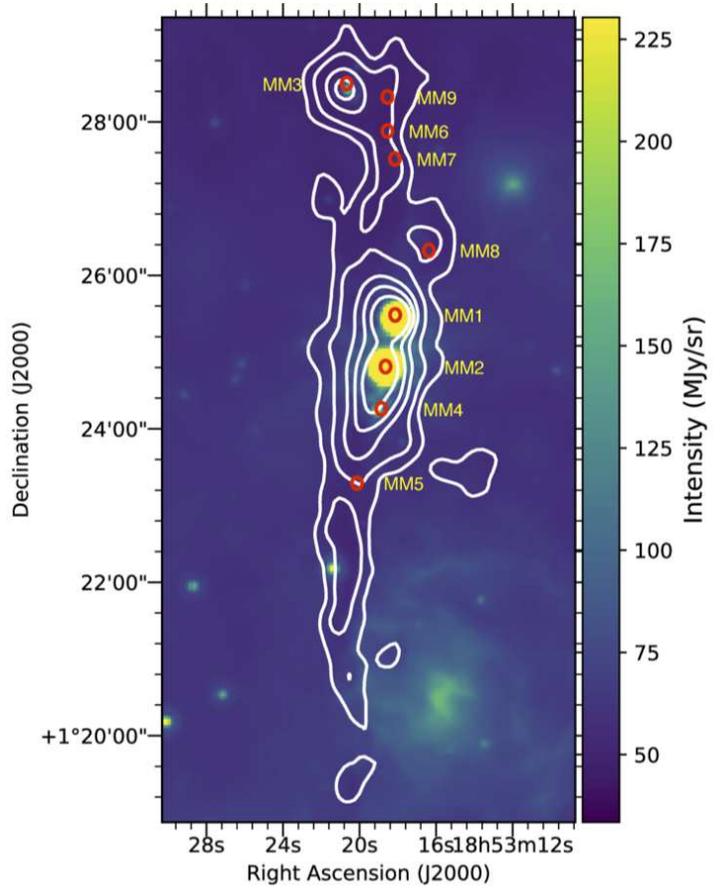}}
\caption{\textit{Spitzer} 24\,$\mu$m image of G34 filament overlaid with JCMT 850\,$\mu$m contours with levels at 45, 144, 418, 800, and 1500 $\rm mJy~beam^{-1}$. The millimeter cores identified by \citet{2006ApJ...641..389R} are shown as red open circles and are labeled as therein (MM1 to MM9).}\label{Fig:spitz24}
\end{figure}

In this work, we report POL-2 observational results toward a more evolved filament, G34.43+0.24 (hereafter G34). At a distance of $\sim$3.7\,kpc \citep{2012ApJ...756...60S, 2014ApJ...791..108F, 2016ApJ...819..117X}, G34 is an active high-mass star-forming filamentary cloud \citep{1998A&A...336..339M, 2011ApJ...741..120R, 2018ApJ...857...35S}. G34 harbors multiple cores, including G34-MM1 through MM9, that are likely at different evolutionary stages \citep{2011ApJ...743..196C}. Figure  \ref{Fig:spitz24} shows the locations of these MM sources. G34-MM2 was found to be the most evolved core \citep{2006ApJ...641..389R} associated with the ultra-compact H\,II (UCH\,II) region IRAS 18507+0121 of spectral type B0.5 \citep{1998A&A...336..339M, 2004ApJ...602..850S, 2007ApJ...669..464S}. Investigating the cores in G34, \citet{2008ApJ...689.1141R} found that the brightest millimeter core, G34-MM1, exhibits a typical chemical signature of a high mass core. On the other hand, the clump MM3 hosts a hot-corino \citep{2014ApJ...794L..10Y, 2015ApJ...803...70S}. \citet{2009ApJS..181..360C} found G34-MM1, MM3, MM4, MM5, and MM8 associated with extended \textit{Spitzer} 4.5\,$\mu$m emissions, indicating possible outflow activities. \citet{2010ApJ...715...18S} also observed these cores and found molecular outflows associated with cores  G34-MM1, MM2, MM3, and MM4. G35 is a filament similar to G34 with several embedded low-luminosity massive protostars \citep{2011A&A...535A..76N} and massive starless clumps \citep{2018ApJ...859..151L}.  A network of filaments covering a broad range of densities is also revealed in G35. The magnetic field lines in G35 tend to be perpendicular to the densest part of the most massive filament, whereas they tend to be parallel in the low-density regions as well as in other less dense filaments \citep{2018ApJ...859..151L}. The magnetic fields together with turbulence, however, do not appear able to support against the gravitational collapse of the densest clumps in G35. The northern region of G34 with MM3 is associated with the PGCC G34.50+0.27. G34 has a mass per unit length of $\sim$1600 M$_{\odot}$~pc$^{-1}$ \citep{2016ApJ...819..117X} which is about four times larger than that ($\sim410$ M$_{\odot}$~pc$^{-1}$) of G35 \citep{2018ApJ...859..151L}. By comparing G34 with G35, we can determine which of the three mechanisms, B-fields, gravity, or turbulence is dominant in filament evolution and dense core formation.


\begin{figure*}
\centering
\resizebox{18.5cm}{10.0cm}{\includegraphics{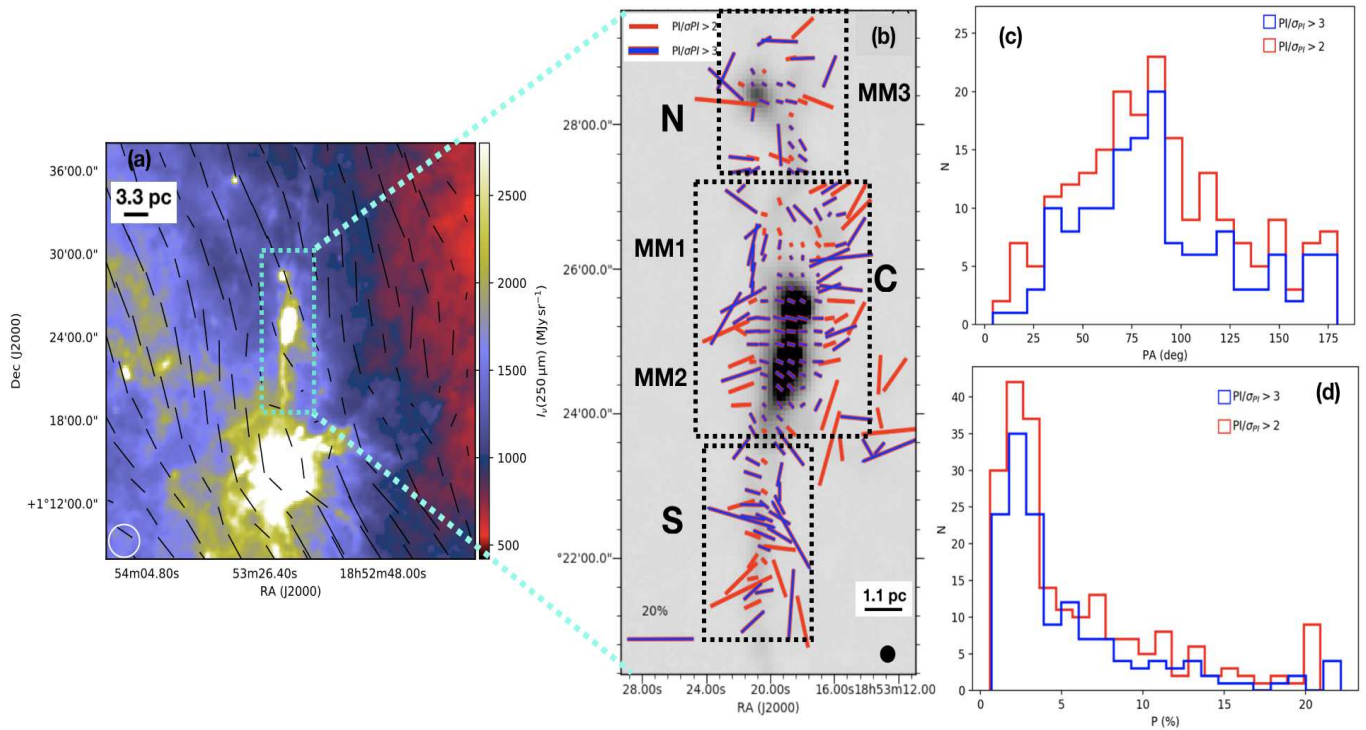}}
\caption{Panel (a) shows the large-scale B-field morphology towards G34 region obtained from Planck 353\,GHz dust polarization observations overlaid on \textit{Herschel} 250\,$\mu$m image. The location of G34 is inside the cyan dashed rectangle in the center. The Planck beam size is shown as an open circle. Panel (b) shows the smoothed (the $\rm 12{\arcsec}$ pixel) B-field orientation in G34 filament from 850\,$\mu$m POL-2 observations. The background greyscale image shows the dust continuum intensity image. Three regions, \enquote*{N}, \enquote*{C}, and \enquote*{S}, are labeled and the JCMT beam size is shown as a solid circle. The vectors correspond to data with $\rm PI/\sigma PI > 2$ (red) and $\rm PI/\sigma PI > 3$ (purple). The scale vector with 20\% polarization is also shown. Panels (c) and (d) are distributions of position angle and polarization fraction for the two datasets, respectively.}\label{Fig:SN_2_3}
\end{figure*}

\section{Observations, data acquisition, reduction and validation} \label{sec:obs}

The POL-2 observations were conducted in August 2018 (M18AP041; PI: Soam A.) in Band-2 weather conditions using the POL-2 daisy map mode of JCMT \citep{2013MNRAS.430.2513H, 2016SPIE.9914E..03F, 2019inpreparation} at 850\,$\mu$m. In this mode of observations, a fully sampled circular region of 11$\arcmin$ diameter is produced with a high signal-to-noise coverage over the central 3$\arcmin$ of the map. This observing mode is based on the SCUBA-2 CV daisy scan pattern \citep{2013MNRAS.430.2513H} but modified to have a slower scan speed (i.e. 8$\arcsec$/s compared to 155$\arcsec$/s) to obtain sufficient on-sky data for good Stokes Q and U values. Coverage decreases, with a consequent significant increase in the rms noise, toward the edges of the map. The POL-2 polarimeter with a rotating half-wave plate at a frequency of 2 Hz and a fixed polarizer is placed in the optical path of the SCUBA-2 camera. The total on-source integration time was $\sim$3 hours with $\tau_{225}$ ranging from 0.05 to 0.08, where $\tau_{225}$ is atmospheric opacity at 225 GHz. We adopted the same observational strategy as described by \citet{2017ApJ...842...66W}.  POL-2 simultaneously collects the data at 450\,$\mu$m and 850\,$\mu$m wavelengths with full-width half maximum (FWHM) of 9.6$\arcsec$ and 14.1$\arcsec$, respectively \citep{2013MNRAS.430.2534D}. We have not reported 450\,$\mu$m data in this work since the instrumental polarization (IP) model for 450\,$\mu$m data was not commissioned when this project started. 

The data were acquired from the Canadian Astronomy Data Center (CADC) and reduced using the STARLINK/SMURF package {\tt\string pol2map} \citep{2013MNRAS.430.2545C, 2014ASPC..485..391C} specifically developed for reducing sub-millimeter data obtained from JCMT. The details of the data reduction steps and procedure are described in \citet{2019ApJ...876...42W}. In the first run of {\tt\string pol2map}, the raw bolometer time-streams for each observation are converted into separate Stokes Q, U, and I time-streams using the process {\tt\string calcqu}. Then a Stokes I map is created from all I time-streams using a routine {\tt\string makemap} which is an iterative map-making process in SMURF package. Individual I maps corresponding to each observations were coadded to produce the initial I map of the region. The details of this step can be seen in \citet{2013MNRAS.430.2545C}. The final I, Q, and U maps were obtained by running {\tt\string pol2map} a third time. The initial I map described in a previous step is used to generate a fixed SNR-based mask for all further iterations of {\tt\string makemap}. The pointing corrections determined in the previous step were applied during the map-making process. During the final process, we invoked an additional parameter called {\tt\string skyloop}\footnote{http://starlink.eao.hawaii.edu/docs/sc22.pdf} in {\tt\string pol2map} and corrected for the loss of synchronization between data values and pointing information in the data reduction process.  {\tt\string Skyloop} improves the recovery of faint, extended structures by performing one iteration of the mapmaker on all of the observations, co-adding the result, and testing each successive iteration for convergence \citep[see][]{2019ApJ...876...42W}. This is in contrast to the traditional map-making method of deriving an iterative solution for each observation individually. The final polarization values obtained here are de-biased by using the mean of Q and U variances to remove statistical bias in regions of low signal-to-noise ratio (SNR).

The calibrated I, Q, and U maps were obtained in $\rm Jy\,beam^{-1}$ units using a Flux Calibration Factor (FCF) of 537 Jy\,$\rm pW^{-1}$ given for 850\,$\mu$m. The output maps are multiplied by 1.35 to account for additional losses due to POL-2 \citep{2013MNRAS.430.2534D, 2016SPIE.9914E..03F}. The final co-added total intensity map has an rms noise\footnote{This value was measured using SCUBA2-MAPSTATS recipe under PICARD package in STARLINK} of $\rm \sim\,7.0\,mJy\,beam^{-1}$. The rms noise in Q and U maps were found to be $\rm \sim\,7.9\,mJy\,beam^{-1}$ and $\rm \sim\,6.8\,mJy\,beam^{-1}$, respectively.

After the final step of running {\tt\string pol2map}, we obtain a polarization vector catalogue which is produced by co-adding Stokes  I, Q, and U maps. The data were reduced with a 4$\arcsec$ pixel size but to improve the sensitivity, we binned the co-added Stokes I, Q and U maps to 12$\arcsec$ pixel size using binning over 3$\times$3 pixels.

The debiased \citep{1974ApJ...194..249W} polarization fraction values were estimated (see \citet{2018ApJ...861...65S, 2019ApJ...876...42W}) as 

 \begin{equation}
 P=\frac{1}{I}\sqrt{Q^{2}+U^{2}-\frac{1}{2}(\delta Q^{2}+\delta U^{2})}   \,\, ,
 \end{equation}
 where P is the debiased polarization fraction and I is the total intensity. Q, U, $\delta Q$, and $\delta U$ are the Stokes parameters and their uncertainties. The uncertainty in polarization fraction is estimated using

\begin{equation}
\delta P = \sqrt{\frac{(Q^2\delta Q^2 + U^2\delta U^2)}{I^2(Q^2+U^2)} + \frac{\delta I^2(Q^2+U^2)}{I^4}}  \,\, .
\end{equation}

The polarization position angles were measured increasing towards the east from the north in the sky projection using relation 

\begin{equation}
{\rm \theta = \frac{1}{2}tan^{-1}(U/Q)}  \,\, .
\end{equation}

The corresponding uncertainties in $\theta$ were calculated using

\begin{equation}
\delta\theta = \frac{1}{2}\frac{\sqrt{Q^2\delta U^2+ U^2\delta Q^2}}{(Q^2+U^2)} \times\frac{180\degree}{\pi}  \,\, .
\end{equation}

The plane-of-sky B-field orientation is inferred by rotating polarization angles by 90$\degree$ (assuming that the polarization is caused by elongated dust grains aligned perpendicular to the magnetic field). We have used only the data points where the observed uncertainties in position angles are less than 20$\degree$. The large-scale B-fields are examined using Planck 850\,$\mu$m (353 GHz) dust emission polarization maps \citep{2015A&A...576A.104P, 2016A&A...586A.138P}. The image is smoothed to the 7$\arcmin$ resolution to ensure good SNR data. The vectors are drawn at 3.5$\arcmin$ (half-resolution) steps.

We checked the quality of our data used for analysis by examining the different SNR values derived from polarization intensity (PI) and its uncertainty ($\sigma_{PI}$). In panel (b) of Figure \ref{Fig:SN_2_3}, the B-fields inferred from SNR$>$ 2 (PI/$\sigma_{PI} > 2$; 211 red vectors) and SNR$>$3 (PI/$\sigma_{PI} > 3$; 146 purple vectors) are generally consistent in the regions where both are available. The other two panels (c) and (d) of Figure \ref{Fig:SN_2_3} show comparisons of the distributions of B-field position angles and polarization fraction of the two subsamples. The aim is to test the validity of the data with $\rm 2<SNR<3$ (which is generally used in such studies) and to decide whether or not data with SNR$\geq$2 could be used for studying B-field morphology and strength. The very similar trends in distributions of position angles and polarization percentages reassures us that we can use the $\rm 2<SNR<3$ data for further analysis.

\section{Results and discussion} \label{sec:optPol}

The Stokes I map of the G34 filament in 850\,$\mu$m continuum emission with inferred B-field geometry is shown in panel (b) of Figure \ref{Fig:SN_2_3}. The elongated shape of the filament is clearly visible and three regions of interest (\enquote*{N}, \enquote*{C}, and \enquote*{S}) are indicated by labeled dashed black rectangles. The overall filament appears to have a small head to the North (containing MM3), a dense clump (consisting of MM1 and MM2) in the center, and a diffuse tail-like structure to the south.

\begin{figure}
\resizebox{8.0cm}{12.0cm}{\includegraphics{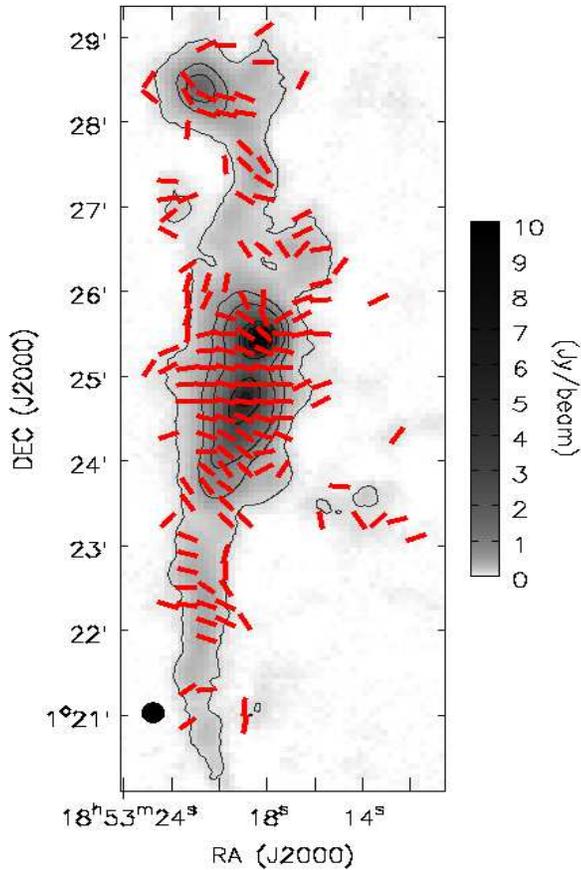}}
\caption{B-field orientation (after 90$\degree$ rotation of the polarization vectors) in G34 shown with normalized line-segments independent of polarization fraction and correspond to $\rm PI/\sigma_{PI} > 3$ and $\rm I/\delta I > 10$ where I and $\delta$I are the total intensity and its uncertainty, respectively. The background image shows the 850$~\mu$m continuum emission overlaid with contours of levels [0.1, 0.5, 1.0, 3.0, 5.0, 7.0, 9.0]\,$\rm Jy\,beam^{-1}$. JCMT beam size is shown with a black solid circle.}\label{Fig:pol2smooth}
\end{figure}

\begin{figure} 
\resizebox{9.6cm}{8.0cm}{\includegraphics{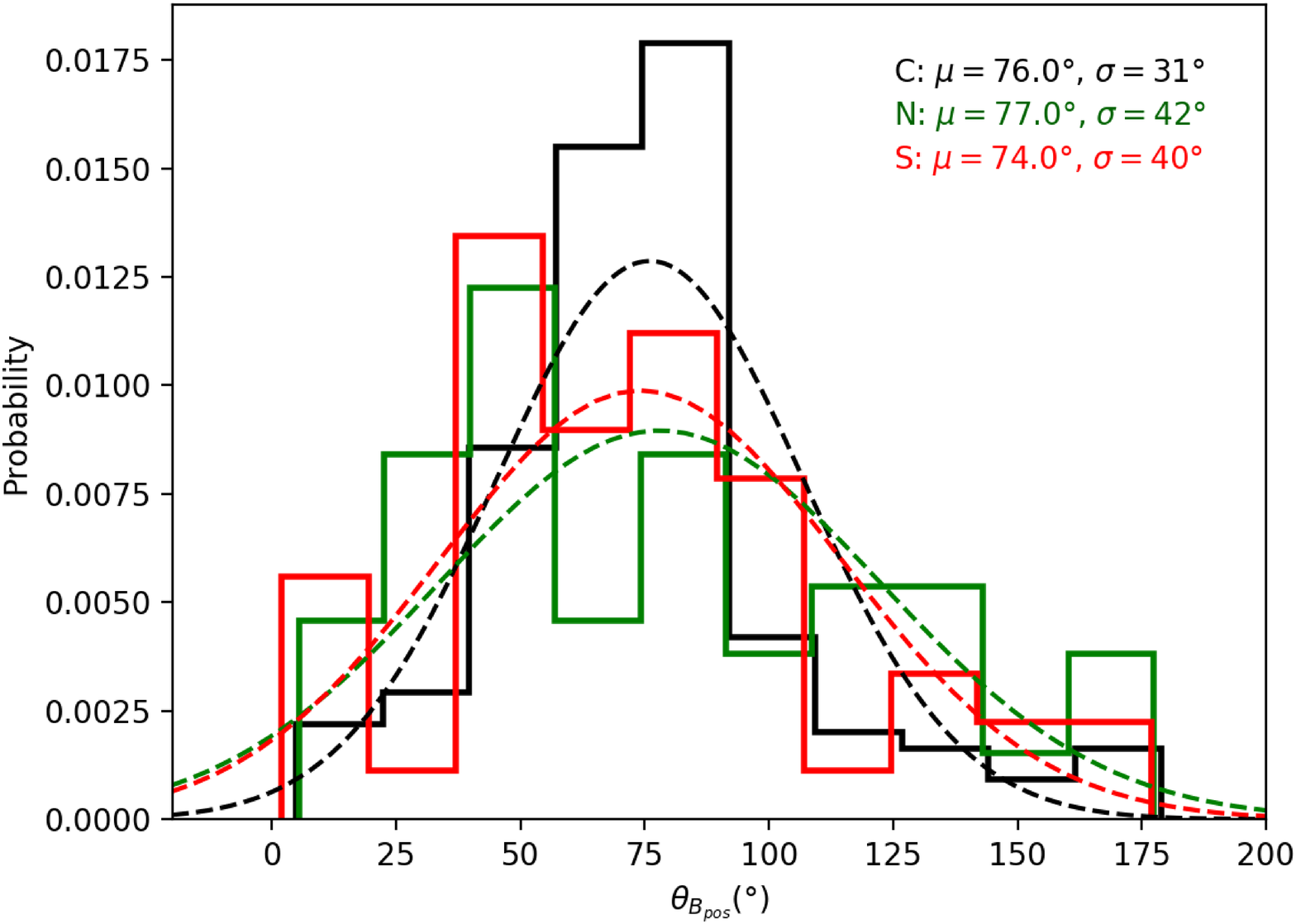}}    
\caption{Gaussian fitted histograms of the B-field position angles of data with PI/$\sigma_{PI} > 2$ in the north (green), south (red), and center (black) of G34.}\label{Fig:histCNS}
\end{figure}

\subsection{Magnetic field morphology} \label{sec:NIRpol}

Panel (a) in Figure \ref{Fig:SN_2_3} represents the large-scale B-fields inferred from Planck measurements in the region containing G34 \citep{2015A&A...576A.104P}. There is a clear indication of field lines aligned in the south-west to north-east directions. Panel (b) of the figure shows zoomed-in B-fields in G34 from our POL-2 observations at sub-parsec scales. The lengths of line-segment are proportional to the fractional polarization. Magnetic field geometry and properties are studied individually in the regions center (C), north (N), and south (S) labelled in panel (b). The northern part containing MM3, has field orientations almost along the elongated clump. The central region, however,  has field lines perpendicular to the long axis of the filament with an apparent smooth change in orientation when seen from west to east. The southern diffuse region has most field lines closely perpendicular to the tail. The large-scale field in the northern region from Planck observations is also nearly parallel to the filament (see left panel (a)), which is similar to the fields seen in the region \enquote*{N} from POL-2 observations. This suggests that the B-field is connected from parsec to sub-parsec scales, despite orders of magnitude difference in density and the physical scales. However, it should also be noted that compared to G34, the region measured by Planck next to it is mostly background and foreground. Hence, it is not evident a priori that the field orientations are identical. We found that the background subtraction would tend to make Planck polarization vectors more perpendicular to the filament but details depend on the selection of the reference regions chosen to represent the background and the filament remains unresolved in the Planck data.

Figure \ref{Fig:pol2smooth} shows a better view of magnetic field morphology in the G34 filament where we use the normalized vectors with their lengths independent of the polarization fraction. The smooth change in field lines from being perpendicular to almost parallel from the center to north regions can be clearly seen in this figure.

Figure \ref{Fig:histCNS} shows the Gaussian fitted distributions of B-field position angles in the center, north, and south regions. The distributions in all regions peak around $75\degree$, which is close to an east-west orientation. 

\subsection{Dust temperature and column density}

\begin{figure*}
\centering
\resizebox{15.0cm}{10cm}{\includegraphics{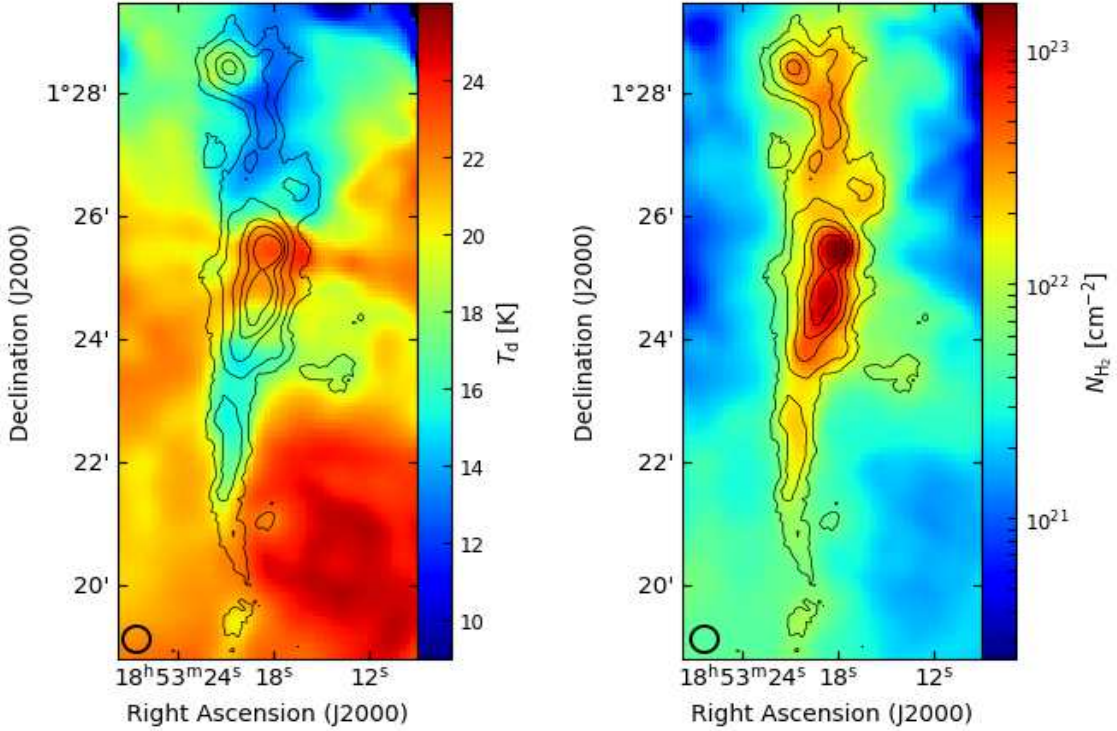}}    
\caption{Left and right panels show the dust temperature and $\rm H_{2}$ column density maps of G34, respectively. The overlaid contours represents 850\,$\micron$ emission.}\label{Fig:cd}
\end{figure*}

We estimated the dust temperature ($T_\mathrm{d}$) and $\mathrm{H_2}$ column density ($N_\mathrm{H_2}$) of the G34 filament using archival \textit{Herschel} PACS/SPIRE (70, 160, 250, 350, and 500 \micron) and JCMT 850 \micron\ data fitted with a modified black-body function. In this procedure, the different resolution \textit{Herschel} images and JCMT 850 \micron\ image were smoothed to the SPIRE 500 \micron\ FWHM beam size of 35\arcsec and reprojected on the same grid. The G34 filament is found embedded in a large-scale molecular cloud in \textit{Herschel} images causing additional emission from surrounding material in the line-of-sight. In order to obtain an accurate column density for G34, this background was subtracted. Then the spectral energy distribution (SED) was fitted to the fluxes obtained in $\textit{Herschel}$ and JCMT maps for each pixel position using the formulae \citep{2008A&A...487..993K}.

\begin{equation}
    I_\nu = B_\nu (T_\mathrm{d})(1-\mathrm{e}^{-\tau_\nu}) \;,
\end{equation}
\begin{equation}
    B_\nu(T_\mathrm{d}) = \frac{2h\nu^3}{c^2}\frac{1}{\mathrm{e}^{h\nu/k_\mathrm{B}T_\mathrm{d}}-1} \;,
\end{equation}
\begin{equation}
    \tau_\nu = \mu_\mathrm{H_2} m_\mathrm{H} \kappa_\nu N_\mathrm{H_2} \;,
\end{equation}
where $B_\nu(T_\mathrm{d})$ is the Planck function at a given dust temperature ($T_\mathrm{d}$), $\tau_\nu$ is the optical depth, $\mu_\mathrm{H_2}$ is the mean molecular weight per hydrogen molecule, $m_\mathrm{H}$ is the hydrogen atom mass, $\kappa_\nu$ is the dust opacity (absorption coefficient), and $N_\mathrm{H_2}$ is $\mathrm{H_2}$ column density. The value of $\mu_\mathrm{H_2}$ is 2.8 and $\kappa_\nu$ for each used frequency are 1.76, 0.4, 0.195, 0.1, 0.05, and 0.0197 cm$^{2}$ g$^{-1}$, respectively, adopted from \citet{1994A&A...291..943O} for a dust-to-gas ratio of 0.01. The temperature and column density maps of G34 made using this procedure are shown in Figure \ref{Fig:cd}. The temperatures (left panel of Figure \ref{Fig:cd}) throughout the filament vary from $\sim$10$-$25\,K with hot dust present in the central region containing MM1/MM2 and colder in the north and southern regions. The column density values (right panel of Figure \ref{Fig:cd}) are found peaking at $\sim 10^{23}\,cm^{-2}$ in the central region. 

We also estimated the $\rm H_{2}$ volume densities of the three regions of G34 north, center, and south assuming them to have cylindrical geometry and adopting the procedure explained in section 3.2 of \citet{2018ApJ...859..151L}. The projected lengths (L) of the cylinders corresponding to \enquote*{N},  \enquote*{C}, and \enquote*{S} regions of G34 shown in middle panel of Figure \ref{Fig:SN_2_3} are 1.9, 3.0, and 2.6\,pc, respectively. The mean values of the projected radius (r) of circular ends of these cylinders are measured to be 1.8, 2.2, and 1.1\,pc, respectively. We used these values to estimate volumes of the cylinders and their number densities. The estimated values of volume densities are shown in Table \ref{tab:values}. 

The estimated values of column and volume densities are used in sections \ref{sec:strength} and \ref{sec:critic} for further calculations.

\subsection{Magnetic Field, Gravity, and Turbulence in G34}

\subsubsection{Magnetic field strength} \label{sec:strength}


We estimate the plane-of-sky B-field ($\rm B_{pos}$) strengths in the central dense region \enquote*{C}, north \enquote*{N}, and south \enquote*{S}  regions of G34 using the Davis-Chandrasekhar-Fermi relation (DCF; Davis 1951; Chandrasekhar \& Fermi 1953). The DCF relation assumes a regular field geometry with dispersion indicating a measure of the distortion in the field geometry caused by turbulence. The vector distribution is considered to be Gaussian with a well-characterized standard deviation.
The DCF method is represented by the expression

\begin{equation}\label{eq:dcf1}
B_{pos} = Q\sqrt{4\pi \rho}\frac{\sigma_{v}}{\sigma_{\theta}}\,\, , 
\end{equation}
where $\rho$ is the gas density, $\sigma_{v}$ is the observed velocity dispersion of the gas, and $\sigma_{\theta}$ is the dispersion in polarization angle. The DCF field model assumes that $Q$ is a factor of order unity that accounts for variations in the B-field on scales smaller than the beam. \citet{2001ApJ...546..980O} compared their mean values of the known plane-of-the-sky magnetic fields with DCF estimates  and found $Q$ in the range of 0.46-0.51. They suggested that the DCF estimate, modified by a multiplicative factor of $\sim$0.5 to account for a more complex magnetic field and density structure, can provide an accurate value of B-field strength when polarization angles are quite uniform. Therefore, we adopted $Q$ as 0.5 for our calculations. Following the simplification introduced by \citet{2004ApJ...600..279C}, eq. \ref{eq:dcf1} can be written as
\begin{equation}\label{eq:dcf}
B_{pos} \approx 9.3\sqrt{n({\rm H}_{2})}\frac{\Delta v}{\sigma_{\theta}}\,\mu{\rm G} \,\, ,
\end{equation}
where $n({\rm H}_{2})$ is the number density of molecular hydrogen in cm$^{-3}$, $\Delta v$  = $\sigma_{v}\sqrt{8\ln 2}$ is FWHM in km\,s$^{-1}$ and $\sigma_{\theta}$ is in degrees. 

The uncertainty in field strength is measured by combining uncertainties using the relation
\begin{equation}\label{eq:dcferr}
 \frac{\delta B_{pos}}{B_{pos}} = \frac{1}{2}\frac{\delta n({\rm H}_{2})}{n({\rm H}_{2})} + \frac{\delta \Delta v}{\Delta v} + \frac{\delta\sigma_{\theta}}{\sigma_{\theta}} \,\, ,
\end{equation}
where $\delta n({\rm H}_{2})$, $\delta \Delta v$, and $\delta\sigma_{\theta}$ are the uncertainties in $n({\rm H}_{2})$, $\Delta v$, and $\sigma_{\theta}$, respectively.

\begin{figure}
\resizebox{8.6cm}{8.0cm}{\includegraphics{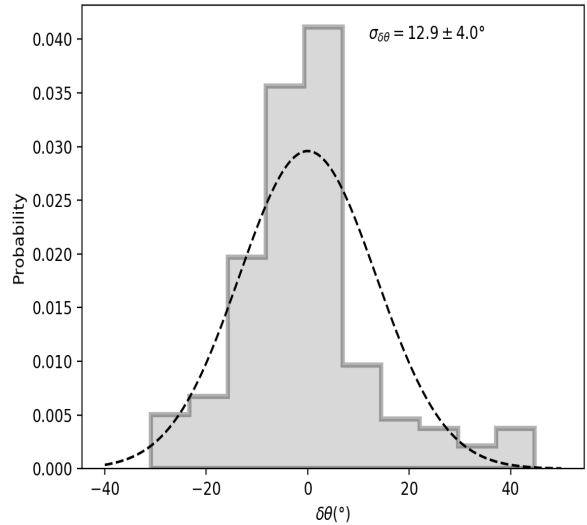}}     
\caption{Distribution of residual position angles ($\rm \delta\theta$ with PI/$\sigma_{PI} > 3$) in the central region of G34.}\label{Fig:hist_residual}
\end{figure}

 \citet{2019ApJ...878...10T} has recently investigated the magnetic field strengths in MM1, MM2, and MM3 regions using CSO/SHARP polarization data at 350\,$\mu$m wavelength and $\rm N_{2}H^{+}$(J=1-0) line observations. We used their velocity dispersion in $\rm N_{2}H^{+}$(J=1-0) line with FWHM $\Delta v$= 1.1$\pm$0.1 $\rm~km~s^{-1}$. We estimated the average volume densities as $\rm \sim 1.8\times10^{5}~cm^{-3}$ in the central region containing MM1/MM2, $\rm \sim 0.6\times10^{5}~cm^{-3}$, and $\rm \sim 0.2\times10^{5}~cm^{-3}$ in the north and south regions, respectively. We subtracted the mean value of position angles from all measured position angles (with $\rm PI/\sigma PI > 3$) in region \enquote*{C} giving residual angles ($\rm \delta \theta$). The Gaussian fit to the distribution of $\rm \delta \theta$ values provides a dispersion in position angle ($\rm \sigma_{\theta}$) of $\rm 12.9\pm4.0\degree$ (see Figure \ref{Fig:hist_residual}). We corrected the value of $\rm \sigma_{\theta}$ by mean value of observed position angle uncertainties which is measured to be $\rm 7.0 \degree$. The corrected value of $\rm \sigma_{\theta}$ is $\rm \sqrt{12.9^{2} - 7.0^{2}} \approx 10.8 \degree$. This value of dispersion in position angles satisfies one of the assumptions of the DCF relation which limits the maximum value of $\rm \sigma_{\theta}$ to be $\rm \leq 25 \degree$ \citep{2001ApJ...561..800H}. Using the above-mentioned values of FWHM in $\rm N_{2}H^{+}$(J=1-0), $n({\rm H}_{2})$, and $\rm \sigma_{\theta}$, the strength of $\rm B_{pos}$ in G34 center is found to be 470$\pm$190 $\mu$G. This field strength is similar to that found in other IRDCs such as $\sim$270\,$\mu$G in G11.11-0.12 \citep{2015ApJ...799...74P} and $\sim$100\,$\mu$G in G035.39-00.33 \citep{2018ApJ...859..151L} but smaller than the value 790\,$\mu$G in G9.62+0.19 \citep{2018ApJ...869L...5L}.

We have used a similar approach as described above to estimate the magnetic field strengths in the North and the South regions of G34. The values of velocity dispersion i.e. FWHM of $\rm N_{2}H^{+}$(J=1-0) in these regions are adopted from \citet{2019ApJ...878...10T} for estimating the field strength using eq. \ref{eq:dcf}. Dispersion in position angles towards these regions is estimated similarly as described above for the central region. The uncertainties in $B_{pos}$ are estimated using eq. \ref{eq:dcferr} which is derived from eq. \ref{eq:dcf} by propagating the errors in the quantities. We found the values of $\rm B_{pos}$ as 100$\pm$40 $\mu$G and 60$\pm$34 $\mu$G in \enquote*{N}, and \enquote*{S} regions, respectively.

\subsubsection{Mass-to-flux ratio} \label{sec:critic}

We will use our estimates of magnetic field strength to calculate the standard parameters of mass-to-flux ratio ($\rm M/\phi_{B}$) and Alfv\'{e}nic Mach number ($\rm M_{A}$). These measure the relative importance of magnetic fields versus gravity and turbulence, respectively.

$\rm M/\phi_{B}$ is the ratio of the mass (M) of the object to the flux ($\rm \phi_{B}$) of the magnetic fields threading the object. \citet{2004ApJ...600..279C} (and references therein) discussed that the maximum mass that can be supported by a given magnetic flux is known as \textit{critical mass}, $\rm M_{crit} = \frac{\phi_{B}}{2\pi \sqrt{G}}$. We tested the importance of the B-field in the context of gravity in all three regions of G34 where plane-of-sky B-field strength is estimated using eq. \ref{eq:dcf}. This can be investigated by calculating the value of criticality parameter ($\rm \lambda_{obs}$) using the relation 

\begin{equation}
\lambda_{obs} = \frac{(M/\phi)_{obs}}{(M/\phi)_{crit}} \,\, ,
\end{equation}
where the observed mass-to-flux ratio is estimated as

\begin{equation}
(M/\phi)_{obs} = \frac{\mu m_{H}N_\mathrm{H_2}}{B_{pos}} \,\, ,
\end{equation}
and $\rm \mu$, $\rm m_{H}$ and $N_\mathrm{H_2}$ are the mean molecular weight per $\rm H_{2}$ molecule, mass of atomic hydrogen, and molecular hydrogen column density, respectively. The average values of column densities in the center, north, and south regions are found to be $\rm \sim 15\times10^{22}$, $\rm 4.5\times10^{22}$, and $\rm 2.5\times10^{22}~cm^{-2}$, respectively.

 The clouds that are not collapsing due to the support by magnetic fields are called magnetically \enquote{subcritical} ($\rm \lambda < 1$), whereas those with gravity which overcomes the support of the magnetic field are referred as magnetically \enquote{supercritical} ($\rm \lambda >1$).


\begin{equation}
(M/\phi)_{crit} = \frac{1}{2\pi \sqrt{G}} \,\, .
\end{equation}

Using the column density in $\rm cm^{-2}$ and measured B-field strength in $\rm \mu G$, we estimated the value of $\lambda_{corr}$ after applying a geometric correction to $\lambda_{obs}$ following \citet{2004Ap&SS.292..225C}. The value of $\lambda_{obs}$ can be overestimated by a factor of 3 due to geometrical effects suggesting $\rm \lambda_{corr} = \lambda_{obs}/3$. The errors in $\lambda_{obs}$ come mainly from the uncertainty in B-field strength. We calculated the $\lambda_{corr}$ in all three regions of G34 and the results are given in Table \ref{tab:values}. The values of $\rm \lambda_{corr}$ obtained for the center, north, and south show that these regions are transcritical. All the values are close to criticality (i.e. $\lambda = 1$) suggesting that gravity and magnetic fields are equally important in these regions.

\subsubsection{B-fields and turbulence}
The nature of turbulent motions in the G34 clump can be studied by estimating the value of $\rm M_{A}$ which describes the relative importance of magnetic fields and turbulence in molecular clouds \citep{2001ApJ...559.1005P, 2008ApJ...687..354N}. When the fields are uniform and strong, the turbulence is regulated by the magnetic fields, yielding a sub-Alfv\'{e}nic scenario (with $\rm M_{A} \leqslant 1$). On the other hand, if the cloud is super-Alfv\'{e}nic (i.e., $\rm M_{A} > 1$), the magnetic field is not strong enough to resist scrambling by turbulent motions. The value of $\rm M_{A}$, using molecular line and polarization observations, can be estimated as

\begin{equation}
M_{A} =   \frac{\sqrt{3}\sigma_{v}}{\sigma_{A}} \,\, ,
\end{equation}
where $\rm \sigma_{v}$ is the mean non-thermal velocity dispersion, measured from the FWHM (i.e. $\rm \sigma_{v} = FWHM/\sqrt{8ln2}$) of $\rm N_{2}H^{+}$(J=1-0) line observations \citet{2019ApJ...878...10T} which we used in estimating magnetic field strength. $\rm \sigma_{A}$ is the Alfv\'{e}nic velocity calculated as

\begin{equation}
\sigma_{A} = \frac{B_{tot}}{\sqrt{{4\pi\rho}}} \,\, ,
\end{equation}

\citet{2004ApJ...600..279C} found from a statistical study that the total magnetic field strength ($\rm B_{tot}$) is 1.3 times the plane-of-sky field strength. In the absence of knowledge of the 3D geometry of G34, this is a reasonable correction to apply. The value of $\rm \sigma_{A}$ in three regions of G34 is calculated using different magnetic field strengths and volume densities in these regions. Using the values of $\rm \sigma_{v}$ and $\rm \sigma_{A}$, we calculate $\rm M_{A}$ in all three regions of G34.

The value of dispersion in position angle, FWHM of $\rm N_{2}H^{+}$(J=1-0) line, volume density, estimated plane-of-the-sky magnetic field strength ($\rm B_{pos}$), projection corrected mass-to-flux ratio ($\lambda_{corr}$), and Alfv\'{e}nic Mach number ($\rm M_{A}$) in regions  \enquote*{C},  \enquote*{N}, and \enquote*{S} are given in Table \ref{tab:values}. The values of $\lambda_{corr}$ in all three regions of G34 suggest it to be marginally critical. The values of Alfv\'{e}nic Mach number suggest the sub-Alfv\'{e}nic nature of turbulence in G34 filament.

Some analytical studies investigated the stability and fragmentation of filaments in context of turbulent motions \citep{1964ApJ...140.1056O, 1992ApJ...388..392I, 2012ApJ...744..190T, 2013ApJ...769..115H} and B-fields \citep{2013ApJ...769..115H} and found B-fields are important in filament formation. \citet{2013ApJ...774..128S} studied the dependence of B-fields on the initial magnetization of filament using combination of synthetic polarization maps and numerical simulations of magnetized clouds and concluded that strong compression is caused by super-Alfv\'{e}nic turbulence. Whereas, the sub-Alfv\'{e}nic turbulence allows the gravitationally collapsing material to move along the B-field lines \citep{2008ApJ...687..354N}. In case of G34, turbulence is found to be sub-Alfv\'{e}nic in all three regions of G34 and the field lines found to be perpendicular to the elongated axes. This is mostly true in the central and southern regions whereas in the northern region the field appears to change from perpendicular to parallel. However, the field orientation seem mostly perpendicular in the lower part of the northern region. \citet{2019ApJ...878...10T} found the similar geometry in G34 north part near MM3 (see Figure \ref{Fig:spitz24} for the location of MM3) and suggest that B-fields must be playing different role here than in central region. \citet{2018ApJ...859..151L} notices field lines getting parallel from perpendicular in northern region of an IRDC G35 suggesting that fields in that region are likely to be poloidal. Similar trend in north of G34 around MM3 agrees with the finding of \citet{2019ApJ...878...10T} and can also indicates that magnetic fields in this region may be poloidal.

\begin{table}
\centering
\caption{Values calculated in G34 Center, North, and South regions.}\label{tab:values}
\scriptsize
\begin{tabular}{llllllc}\hline
Region  & $\rm \sigma_{\theta}$ & $\rm \Delta v$& $\rm n_{H_{2}}$& $\rm B_{pos}$ & $\rm \lambda_{corr}$ & $\rm M_{A}$  \\
        & ($\degree$) & ($\rm km~s^{-1}$)& ($\rm cm^{-3}$)& ($\mu$G) &  & \\\hline
Center    &11$\pm$4 &1.1$\pm$0.1 &$\rm 1.8\cdot10^{5}$&470$\pm$190  & 0.8$\pm$0.4 & 0.34$\pm$0.13\\
North     &16$\pm$9 &0.8$\pm$0.2 &$\rm 0.6\cdot10^{5}$&100$\pm$40  & 1.1$\pm$0.8 & 0.53$\pm$0.30\\
South     &15$\pm$8 &0.6$\pm$0.2 &$\rm 0.2\cdot10^{5}$&60 $\pm$34  & 0.9$\pm$0.5 & 0.49$\pm$0.26\\\hline
\end{tabular}
\end{table}



\subsubsection{Structure function and auto-correlation function analysis}


\begin{table*}
\begin{center}
\caption{{\bf Parameters derived from modified DCF methods without and with correction for beam integration}.}\label{tab:SFACF}
\scriptsize
\begin{tabular}{lllllc}\hline
                                                                                          &                                              & {\bf Without correction}  &  & {\bf With correction} \\
                                                                                          &                                              & {\bf for beam integration} & & {\bf for beam integration}                 \\
Paramters                                                                            & Description                                                    & SF                           & ACF                         & SF                           & ACF \\\hline
$\Delta \theta (^{o})$                                                        &Angular dispersion                                            &$\rm 26.0\pm0.3$   & $\rm 23.2\pm2.7$ &$\rm 52.0\pm2.8$     & $\rm 54.0\pm5.1$\\
$\langle \delta B^{2}\rangle / \langle {B_{0}}^{2}\rangle $ &Turbulent-to-ordered magnetic field energy ratio &$\rm 0.21\pm0.02$&$\rm 0.16\pm0.01$ &$\rm 0.90\pm0.04$  &$\rm 0.81\pm0.01$ \\
$B_{pos}$ ($\mu$G)                                                             &Plane-of-sky magnetic field strength                      &$\rm 300\pm120$  &$\rm 200\pm70$    &$\rm 150\pm90$   &$\rm 90\pm50$\\\hline
\end{tabular}
\end{center}
\end{table*}


 \begin{figure}  
\resizebox{8.5cm}{7.8cm}{\includegraphics{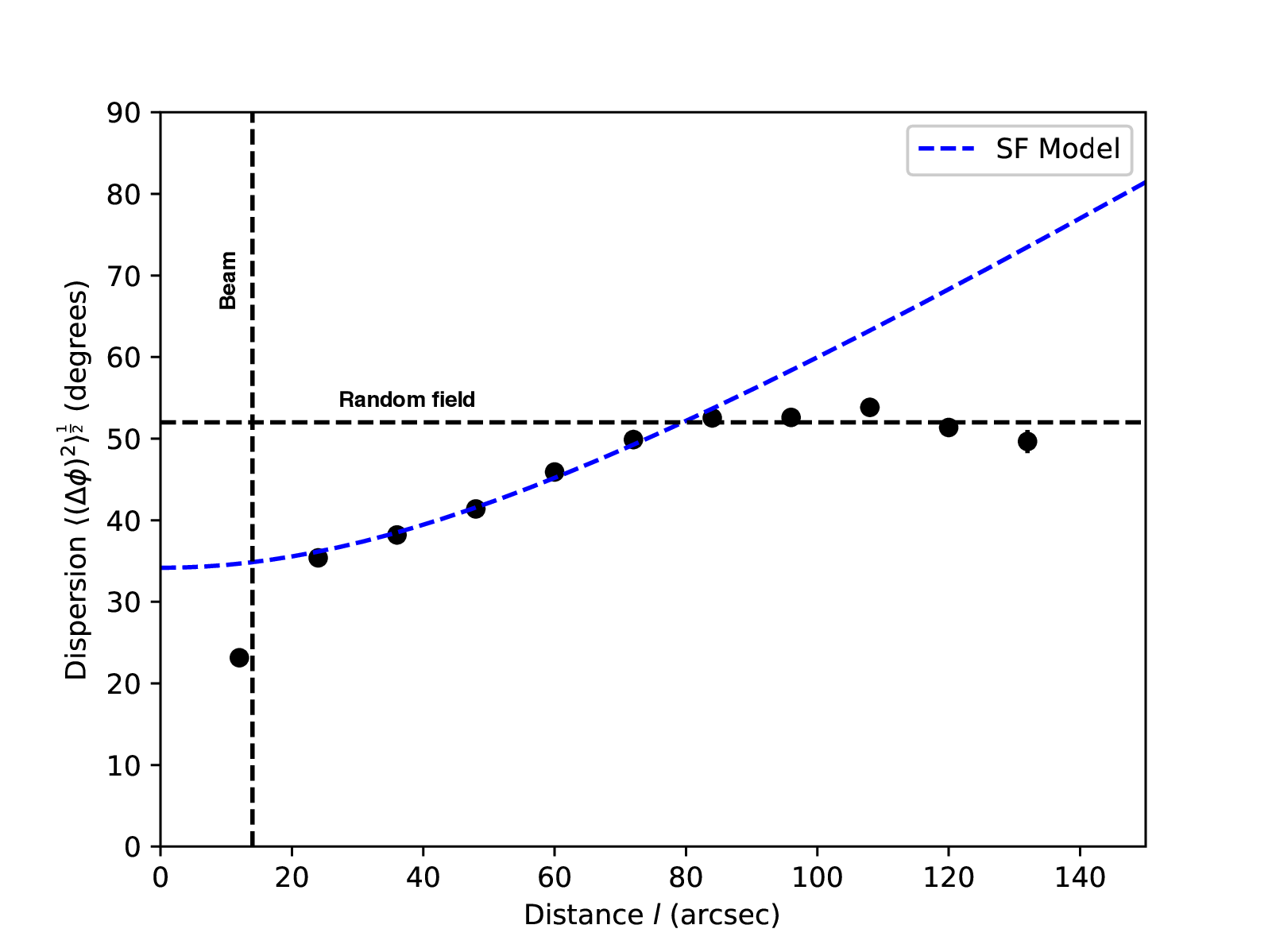}} 
\caption{Angular dispersion function of G34 central region with angle dispersion segments shown with black solid circles and associated error-bars. The best fit is shown with blue dashed line. Vertical dashed line indicates the JCMT beamsize of $14{\arcsec}$ and horizontal dashed line shows the value of angular dispersion function expected for a random field ($52{\degree}$, \citealt{2010ApJ...716..893P}).}\label{Fig:SF}
\end{figure}

We attempted to separate the large-scale and the turbulent scale B-fields in the cloud. In the structure function (SF) method of \citet{2009ApJ...696..567H}, the magnetic field consists of large-scale structure, $B_{0}$ and a turbulent component, $\delta B$. The SF analysis provides the variation of angular dispersion of position angles obtained from polarization observations as a function of separation length $\ell$. The turbulent component $\delta B$ reaches the maximum at some scale larger than the turbulent-scale $\delta$. At scales smaller than $\textit{d}$ (where \textit{d} is the correlation length scale which characterizes the variation in $B_{0}$ \citep{2009ApJ...696..567H}), the higher-order terms in a Taylor expansion of regular component $B_{0}$ can be ignored. In case of $\rm \delta < \ell << \textit{d}$, the angular dispersion function can be written as:
 
 \begin{equation}
 {\langle \Delta \phi^{2}(l)\rangle}_{tot} \simeq b^{2} +m^{2}l^{2} + {\sigma_{M}}^{2}(l) \,\, ,
 \end{equation}
 where ${\langle \Delta \phi^{2}(l)\rangle}_{tot}$ is the square of the total measured dispersion function, where $b^{2}$ is a constant turbulent contribution, $m^{2}l^{2}$ is the contribution from the large-scale field structure, and ${\sigma_{M}}^{2}(l)$ is the contribution of the measured uncertainty. The ratio of the turbulent to large-scale magnetic field components given by 
 
\begin{equation}
\frac{{\langle\delta B^{2}\rangle}^{1/2}}{B_{0}} = \frac{b}{\sqrt{2-b^{2}}} \,\, ,
\end{equation}

and $B_{0}$ is estimated as

\begin{equation}
B_{0} \simeq \sqrt{(2-b^{2})4 \pi \mu m_{H} n_{H_{2}}}\frac{\sigma_{v}}{b} \,\, .
\end{equation}

$B_{pos}$ is corrected by using a correction factor Q as

\begin{equation}
B_{pos} = QB_{0} \,\, .
\end{equation}

The value of Q is taken as 0.5. The angular dispersion function (ADF) corrected by uncertainty $({\langle \Delta \phi^{2}(l)\rangle}_{tot}  - {\sigma_{M}}^{2}(l))$ is shown in Figure \ref{Fig:SF} plotted as a function of distance measured in polarization map. We followed \citet{2009ApJ...696..567H} and divided data into separate distance bins with separations corresponding to the pixel size. At the scales of $0{\arcsec}-25{\arcsec}$, the ADF increases steeply, probably due to the contribution from the turbulent field. After $25{\arcsec}$ length, the dispersion function increases with shallower slope which may be a contribution from the large-scale regular magnetic fields. It reaches the maximum at $\sim 90{\arcsec}$, the maximum ADF value seen here is less than $52{\degree}$ the one expected for random field structure \citep{2010ApJ...716..893P}. The structure function is fitted over $25{\arcsec} < l < 90{\arcsec}$. The calculated parameters are given in Table \ref{tab:SFACF}.


\begin{figure}  
\resizebox{8.5cm}{7.8cm}{\includegraphics{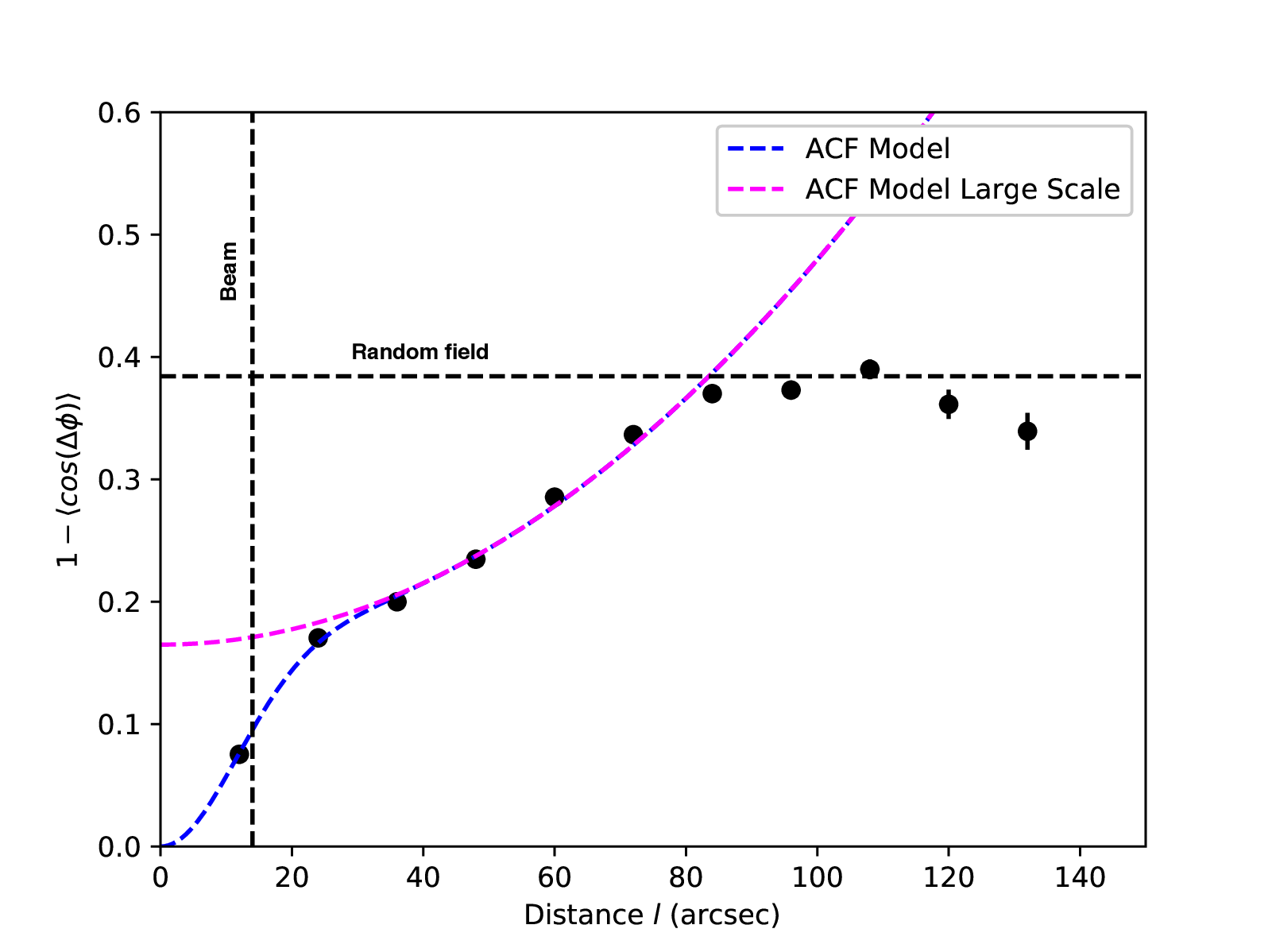}} 
\resizebox{8.5cm}{7.8cm}{\includegraphics{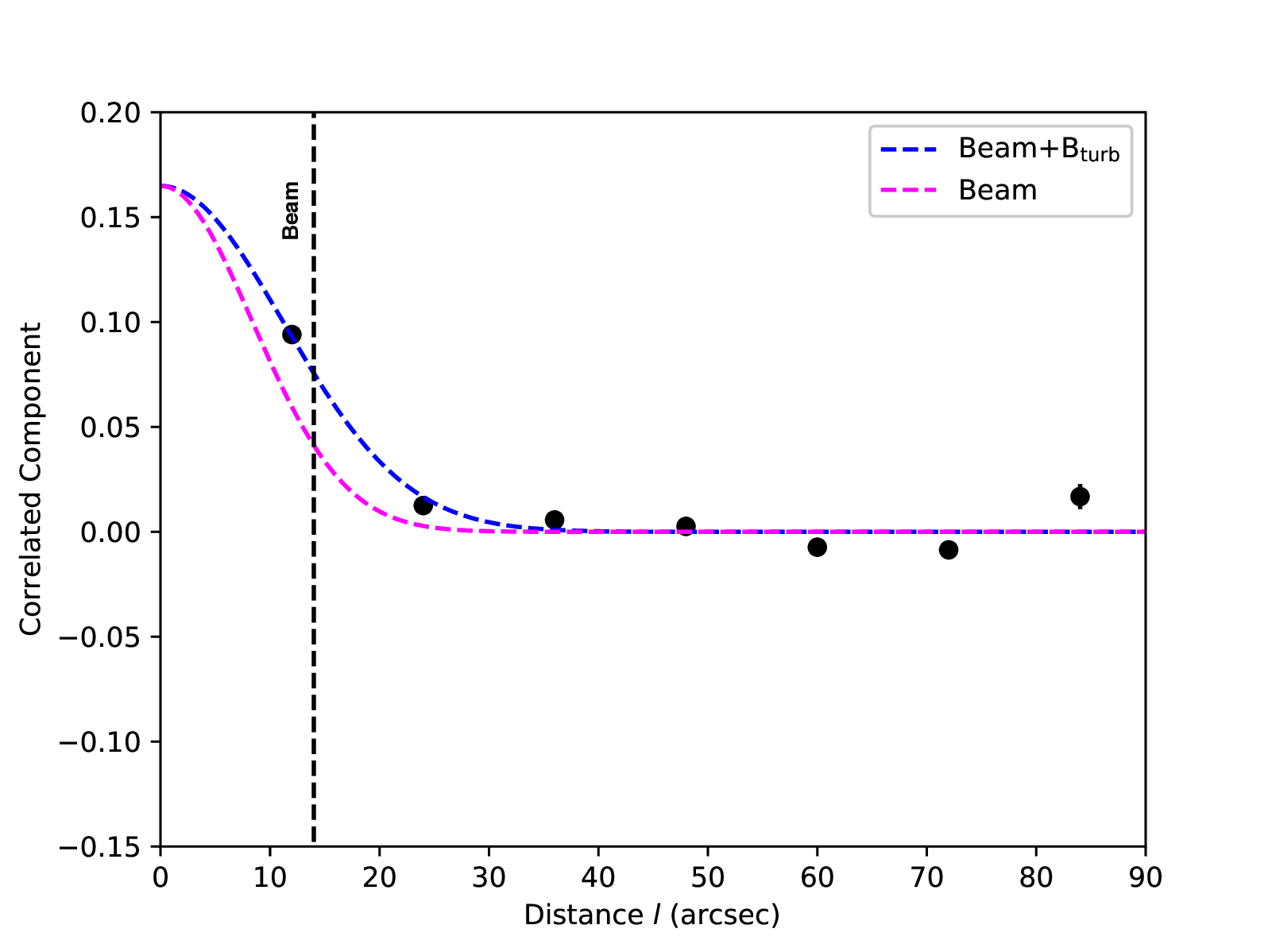}} 
\caption{{\bf Upper panel:} Angular dispersion function for G34 central region with angle dispersion segments shown with black solid circles. The bin size is the same as taken for Figure \ref{Fig:SF}. Blue dashed curve shows the fitted dispersion function. Pink dashed line shows the large-scale component $(1/N)(\langle \delta B^{2}\rangle/\langle B_{0}\rangle) + {a_{2}}^{\prime}l^{2}$ of the best fit. {\bf Lower panel:} Correlated component of the dispersion function $(1/N)(\langle \delta B^{2}\rangle/\langle B_{0} \rangle)e^{-l^{2}/2(\delta^{2}+2W^{2}})$ shown with blue dashed line. The pink dashed line shows the correlated component only due to the beam.}\label{Fig:ACF}
\end{figure}

 The autocorrelation function (ACF) method \citep{2009ApJ...706.1504H} is the expansion of structure function analysis with the inclusion of the effects of signal integration along the line-of-sight and within the beam. The ADF by \citet{2009ApJ...706.1504H} is written as 

\begin{equation}
1-\langle cos[\Delta \phi(l)]\rangle \simeq \frac{1}{N}\frac{\langle\delta B^{2}\rangle}{\langle{B_{0}^{2}}\rangle}\times[1-e^{-l^{2}/2(\delta^{2}+2W^{2})}]+{a_{2}}^{\prime}l^{2} \,\, ,
\end{equation}
where $\Delta\phi(l)$ is the difference between the position angles at a separation of $\ell$, W is the beam radius (6${\arcsec}$ in case of JCMT which is the FWHM beam divided by $\sqrt{8ln2}$), ${a_{2}}^{\prime}$ is the slope of second-order term in the Taylor expansion, and $\delta$ is the turbulent correlation length. N is the number of turbulent cells in telescope beam which is given by

\begin{equation}
N = \frac{(\delta^{2}+2W^{2})\Delta^{\prime}}{\sqrt{2\pi}\delta^{3}} \,\, ,
\end{equation}
where $\Delta^{\prime}$ is the effective thickness of the cloud derived from the distance corresponding to the half-maximum of polarized flux of the cloud \citep{2009ApJ...706.1504H}. The ordered magnetic field strength can be estimated using

\begin{equation}
B_{0} \simeq  \sqrt{4\pi\mu m_{H}n_{H_{2}}}\sigma_{v}[\frac{\langle\delta B^{2}\rangle}{\langle{B_{0}^{2}}\rangle}]^{-1/2} \,\, .
\end{equation}

The upper panel in Figure \ref{Fig:ACF} shows the ADF of polarization segments in G34 \enquote{C} region. The lower panel of the figure shows the correlated component of the dispersion function. The function is fitted at $l < 90{\arcsec}$ distance. The reduced $\chi^{2}$ of the fitting is 5.3. The turbulent correlation length $\delta$ is $\rm 7.4 \pm 0.9{\arcsec}$ (0.13$\pm$0.02\,pc). As mentioned above, the turbulent correlation length characterizes the turbulent component of magnetic fields. This is typical the size of a turbulent magnetized cell. Some previous studies have reported values of the turbulent correlation lengths as $\sim$16\,mpc and $\sim$10\,mpc in the high-mass star forming regions OMC1 \citep{2009ApJ...706.1504H} and Orion KL \citep{2011ApJ...733..109H}. In the starless core Oph-C, \citet{2019ApJ...877...43L} reported correlation length of $\sim$4.3\,mpc. All these regions are much closer compared to G34. The turbulent correlation length in G34 is larger compared to nearby regions due to insufficient power to resolve it at 3.7\,kpc. The number of turbulent cells in G34 is derived as 5.5$\pm$0.3. Other calculated parameters are given in Table \ref{tab:SFACF}. The uncertainties in derived parameters are statistical uncertainties from the dispersion function method. The uncertainty in $B_{pos}$ is taken as a factor of two, as seen in other measurements in several studies. We did not perform this analysis on the northern and southern regions as we do not have enough vectors (20-25 vectors only) for dispersion function analysis and the ADF is too scattered to fit the function. A detailed investigation of change in estimated parameters of SF and ACF analysis on correction with and without beam integration methods can be found in \citet{2019ApJ...877...43L}.


 A detailed comparison of magnetic fields, gravity, and turbulence on filament to core scale in G34 has been presented by \citet{2019ApJ...878...10T} at 350\,$\micron$. In this work, we are comparing these quantities in three different regions of the filament using our POL-2 measurements at 850\,$\micron$.

\subsection{Comparison to other studies}\label{sec:conc}

\begin{figure}
\resizebox{9.0cm}{9.5cm}{\includegraphics{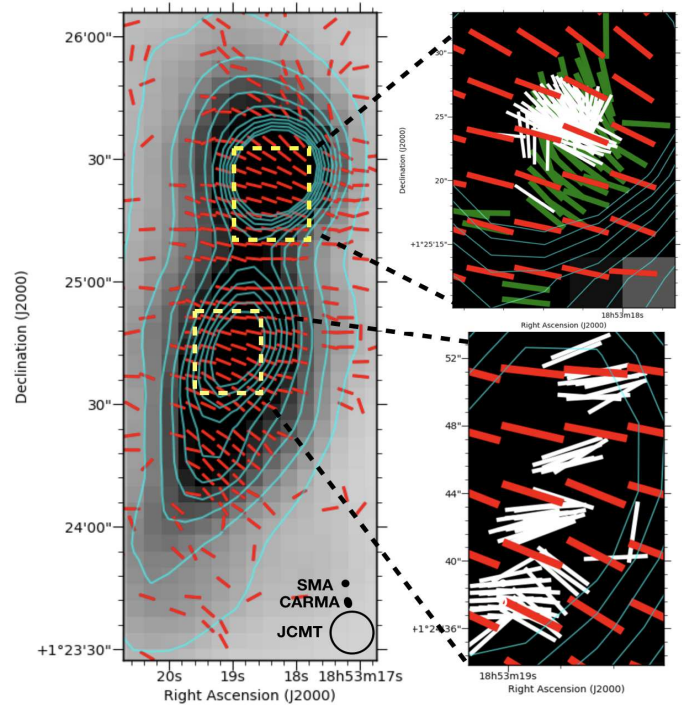}}
\caption{Left panel shows $I_{850{\rm \mu m}}$ (gray map with cyan contours) and the B-field vectors (red lines) in the central clump of G34 obtained from POL-2. The data plotted here correspond to PI/$\sigma_{PI} > 2$. The region of CARMA and SMA observations are marked with the yellow dashed rectangles. Right panel shows the zoomed-in regions with the B-field mapped from CARMA 1.3\,mm observations in the MM1 core (green lines) and from SMA observations at 870\,$\mu$m (white lines). The resolutions of POL-2, CARMA and SMA observations are 14$\arcsec$, 2.5$\arcsec$, and 1.5$\arcsec$, respectively. The labelled beam-sizes are shown in the left panel.}\label{Fig:normalCent}
\end{figure}

There have been several attempts to investigate the B-fields in G34 filament in various wavelengths using dust and line emission polarization measurements. Figure \ref{Fig:normalCent} shows the field morphologies mapped by JCMT/POL-2 (this work), TADPOL/CARMA\footnote{Combined Array for Research in Millimeter-wave Astronomy} \citep{2014ApJS..213...13H}, and the SMA\footnote{Submillimeter Array} \citep{2014ApJ...792..116Z} observations towards the G34 center containing MM1 and MM2. The field orientation seems to be similar from large to small scales when seen from POL-2 (red vectors) and TADPOL observations (green vectors). But the field geometry changes on even smaller scales seen with the SMA (white vectors). The difference in the POL-2 and SMA field geometries can be seen in zoomed-in lower right panel of Figure \ref{Fig:normalCent} where white line segments are misaligned with red lines. A quantitative comparison using histograms of B-field position angles from JCMT, CSO, SMA, and CARMA is shown in Figure \ref{Fig:hist}. The details of these other investigations of B-fields in G34 are given below.

\begin{figure} 
\resizebox{8.5cm}{7.8cm}{\includegraphics{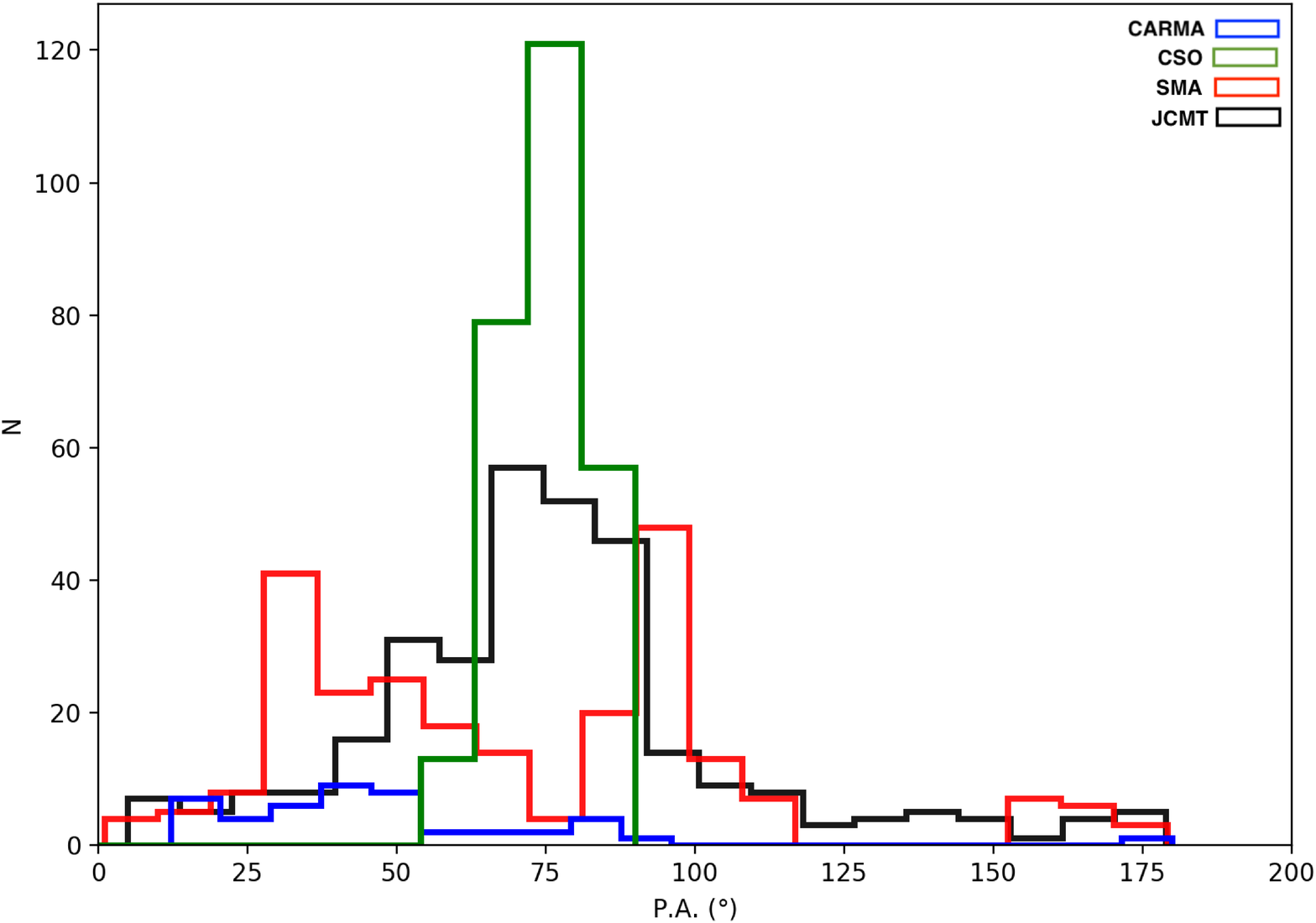}}    
\caption{Histograms of the B-field position angles in G34 central region with JCMT, CARMA, CSO, and SMA observations.}\label{Fig:hist}
\end{figure}

\citet{2019ApJ...878...10T} studied the details of magnetic fields in the regions of G34 containing MM1, MM2, and MM3 using high-resolution (i.e. 10$\arcsec$) 350\,$\mu$m CSO/SHARP polarization observations and kinematics using $\rm N_{2}H^{+}$(1-0) line observations. The B-field orientation found perpendicular to the main axis of filament as also seen in this work suggests that field lines are guiding material towards the filament. They found a close alignment between local velocity gradients derived from $\rm N_{2}H^{+}$(1-0) line and local B-field orientation. Since our 850\,$\mu$m polarization measurements are consistent with 350\,$\mu$m polarization results of \citet{2019ApJ...878...10T}, we expect the similar correlation of local velocity gradients and B-field lines at 850\,$\mu$m. This kind of correlation suggests a coupling of B-fields and gas motion in G34 filament. \citet{2019ApJ...878...10T} also propose varying relative importances of B-fields, gravity, and turbulence in MM1/MM2 and MM3 resulting in different patterns of small scale fragmentation in the clumps at 0.2\,pc scale. The clump containing MM1 shows no fragmentation at all. They found that clumps containing MM2 show an aligned fragmentation and the other clumps with MM3 show a clustered fragmentation. We refer to \citet{2019ApJ...878...10T} for detailed explanation of these findings.

\citet{2008ApJ...676..464C} presented interferometric observations of polarized continuum in 3 mm wavelength (with 16${\arcsec}$ resolution) and line emission using CO (J=1-0) from G34 filament using BIMA\footnote{Berkeley-Illinois-Maryland Association} array. They found a very uniform polarization pattern from both dust and line emission as seen in present work in 850\,$\mu$m and at 350\,$\mu$m by \citet{2019ApJ...878...10T}. This is a remarkable consistency of polarization measurements in different wavelengths tracing different dust grains.

\citet{2014ApJS..213...13H} have studied the B-fields in G34 central region using $\lambda$ 1.3mm TADPOL/CARMA observations of dust polarization with 2.5${\arcsec}$ resolution. The observations from the present work at 850\,$\mu$m, from the CSO at 350\,$\mu$m and BIMA at 3 mm wavelength show uniform and ordered field geometry in G34 central region but the results of \citet{2014ApJS..213...13H} reveal a much more complex polarization pattern with a dragged B-field geometry. They even see a hint of an hourglass morphology in the densest part of the core.

\citet{2014ApJ...792..116Z} investigated the small scale ($\leq 0.1$ pc) B-field structure in G34 center high-density region using SMA at 870\,$\mu$m wavelength with 1.5${\arcsec}$ resolution. Their findings also suggest that the magnetic fields are roughly perpendicular to the major axis of the filament and consistent with those of \citet{2014ApJS..213...13H} in MM1. The SMA polarization measurements are uniform but deviate from our 850\,$\mu$m B-field orientations.


The zoomed-in panels of Figure \ref{Fig:normalCent} show a deviation in field lines. The sub-parsec scale fields are misaligned and even become perpendicular to the large-scale field lines. Similarly in the MM2 region, sub-parsec field lines probed with SMA observations are almost, if not exactly, perpendicular to the large-scale fields seen with JCMT. The field might be strong enough on the clump scale to guide the material along the field lines which eventually get concentrated into cores. The concentration can pinch the B-field lines inside the cores, but does not necessarily lead to complete misalignment with the large-scale field lines. This may be a potential explanation of the change in field geometry from large clump to small core scales. The MHD simulations by \citet{2019MNRAS.485.4509L} also revealed the deviation of core scale magnetic fields from large-scale average field orientations with a deviation as strong as 90$^{\degree}$. They suggested that change may be caused by the gravitational collapse, enhanced turbulence, and the gas flow along the cloud's long axis. 

This can also be explained with numerical simulations \citep{2001ApJ...546..980O, 2008ApJ...687..354N, 2014ApJ...789...37V} showing less disturbed and organized field geometry when B-fields are stronger i.e. $ \beta = (P_{th}/P_{B}) << 1$, where $\beta$ is the ratio of thermal pressure ($P_{th}$) to magnetic pressure ($P_{B}$). To estimate the  $\beta$ values from our observations in G34, we calculated the magnetic pressure ($P_{B} = B^{2}/8\pi$) and thermal pressure ($P_{th} = nkT$), where B is plane-of-the-sky B-field strength, $n$ is the volume density, and T is the gas temperature \citep{2015AJ....150..159D}. The values of $\beta$ are found to be 0.1, 0.2, and 0.6 in the north, central, and southern regions, respectively. \citet{2016MNRAS.459.1803W} investigated 3D MHD simulations to understand the formation of clumps and filaments and to determine the driving processes responsible for filament formation and fragmentation. They explored the range of magnetic field strengths in clouds with $\beta$ varying from 0.1 to 1.0. They found that with no magnetic fields, clumps are found to be forming within the cloud whereas in the case of strong magnetic fields ($\beta$=0.1) these clumps start appearing as filaments. Our findings of $\beta$ values in G34 are consistent to these simulations and suggest that B-fields are playing important role in the formation of this filament.

Outflow patterns of in MM1, MM2, and MM3 of G34 are plotted with SMA polarization measurements in figure 1 of \citet{2014ApJ...792..116Z}. The outflows in MM1 are compact and mostly aligned with the small scale B-fields (as seen in their figure) but the outflows in MM2 are highly complex with red and blue shifted lobes overlapping each other. Therefore, it is hard to check for any correlation in B-fields and outflows in the core MM2. \citet{2010ApJ...715...18S} reported the discovery of outflows in MM3 using CO (J=3-2) line observations. The outflow mass and kinetic energy associated to outflows in MM3 suggest a high-intermediate mass star embedded in the core. The outflow orientation is not indicated in their work so it is not possible to relate the field orientations and outflow direction in MM3. The authors also report a possible association of outflows with the core MM4 in G34 central region. The highly ordered and the uniform field geometry of G34 seen in the above mentioned studies including the present work, suggests that feedbacks from these detected outflows associated to MM1, MM2, MM3, and MM4 are not significantly affecting the field geometry of the region. This may be further investigated on much smaller scales using ALMA polarization capabilities.

Among the cores embedded in G34, MM2 has an associated UCHII region \citep{2004ApJ...602..850S}. We did not see any prominent change in the B-field lines due to the compression by HII region in this core MM2 as seen by \citet{2018ApJ...869L...5L} in an actively high-mass star-forming region G9.62+0.19. The B-field strength in G34 is found to be less strong than in G9.62+0.19. To further investigate the effect of UCHII regions on B-fields in G34-MM2, we have to probe the fields and kinematics at much smaller scales using ALMA observations as done by \citet{2019A&A...626A..36D} in G9.62+0.19.

\section{Conclusion}\label{sec:conc}

We present the plane-of-sky projected magnetic field in G34, obtained using 850$~\mu$m dust polarization observations. We investigated the relative importance of gravity, turbulence, and magnetic fields in G34 at sub-parsec scales. The main findings of the study are as follows:

\begin{enumerate}

\item The overall B-field structure in G34 is ordered and perpendicular to the long axis of the filament. The small-scale field geometry is found connected to the large-scale field lines seen with Planck dust polarization observations. The observed aligned B-fields in G34 are consistent with theoretical models suggesting that B-fields play an important role in guiding the contraction of the cloud driven by gravity.

\item Our measurements of field geometry in G34 using JCMT 850\,$\mu$m wavelength are found consistent with previous studies which inferred field morphology at 350\,$\mu$m (CSO) and 3\,mm (BIMA) wavelengths. However, there is some deviation in the field lines seen at core scale at 870\,$\mu$m (SMA) and 1.3\,mm (CARMA) wavelengths.

\item The present study, combined with several similar studies of other IRDCs, suggests that field lines are mostly perpendicular to the filament major axes but change direction at sub-parsec scales in embedded cores which may be caused by relatively different roles of gravity and B-fields than that on clump scale.

\item We used an updated form of the Davis-Chandrasekhar-Fermi relation to estimate a plane-of-sky magnetic field strength of 470$\pm$190\,$\mu$G, 100$\pm$40\,$\mu$G, and 60$\pm$30\,$\mu$G in the central, northern, and southern regions of G34, respectively. Our results are consistent with those found in several other observations of IRDCs and behavior predicted by theoretical simulations.

\item From the estimation of mass-to-flux ratio, G34 filament  is found to be marginally critical with a criticality parameter $\rm \lambda_{corr}$ of 0.8$\pm$0.4, 1.1$\pm$0.8, and 0.9$\pm$0.5 in the central, northern, and the southern regions, respectively. 

\item The values of Alfv\'{e}nic Mach number in all three regions correspond to a sub-Alfv\'{e}nic nature of turbulence in G34 filament.

\end{enumerate}

The authors thank referee for a constructive report which considerably helped to improve the content of the manuscript. AS and B-GA are supported by National Science Foundation Grant-1715876. TL is supported by KASI and EACOA fellowships. AS acknowledges the support from Korea Astronomy \& Space Science Institute (KASI) for our JCMT observations. MJ, ERM and VMP are supported by Academy of Finland Grant-285769. VMP is also supported by the European Research Council, Advanced Grant No. 320773, and the Spanish MINECO under projects MDM-2014-0369 and AYA2017-88754-P. CWL was supported by National Research Foundation of Korea (NRF-2019R1A2C1010851). AS thanks Simon Coud$\acute{e}$ for a useful discussion during the revision and Piyush Bhardwaj for a critical reading of the draft. This work was carried out in part at the Jet Propulsion Laboratory, operated for NASA by the CalTech. DJ is supported by National Research Council Canada and an NSERC Discovery Grant. WK was supported by Basic Science Research Program through the National Research Foundation of Korea (NRF-2016R1C1B2013642). JCMT is operated by the East Asian Observatory on behalf of National Astronomical Observatory of Japan; Academia Sinica Institute of Astronomy and Astrophysics; the Korea Astronomy and Space Science Institute; the Operation, Maintenance and Upgrading Fund for Astronomical Telescopes and Facility Instruments, budgeted from the Ministry of Finance of China and administrated by the Chinese Academy of Sciences and, the National Key R\&D Program of China (No. 2017YFA0402700). 

\textit{Facility:} James Clerk Maxwell telescope (JCMT)
\textit{Softwares:} Starlink \citep{2014ASPC..485..391C}, Astropy \citep{2013A&A...558A..33A}.



\bibliographystyle{aasjournal}

\begin{thebibliography}{}

\bibitem[{{Andr{\'e}} {et~al.}(2014){Andr{\'e}}, {Di Francesco},
  {Ward-Thompson}, {Inutsuka}, {Pudritz}, \& {Pineda}}]{2014prpl.conf...27A}
{Andr{\'e}}, P., {Di Francesco}, J., {Ward-Thompson}, D., {et~al.} 2014,
  Protostars and Planets VI, 27

\bibitem[{{Astropy Collaboration} {et~al.}(2013){Astropy Collaboration},
  {Robitaille}, {Tollerud}, {Greenfield}, {Droettboom}, {Bray}, {Aldcroft},
  {Davis}, {Ginsburg}, {Price-Whelan}, {Kerzendorf}, {Conley}, {Crighton},
  {Barbary}, {Muna}, {Ferguson}, {Grollier}, {Parikh}, {Nair}, {Unther},
  {Deil}, {Woillez}, {Conseil}, {Kramer}, {Turner}, {Singer}, {Fox}, {Weaver},
  {Zabalza}, {Edwards}, {Azalee Bostroem}, {Burke}, {Casey}, {Crawford},
  {Dencheva}, {Ely}, {Jenness}, {Labrie}, {Lim}, {Pierfederici}, {Pontzen},
  {Ptak}, {Refsdal}, {Servillat}, \& {Streicher}}]{2013A&A...558A..33A}
{Astropy Collaboration}, {Robitaille}, T.~P., {Tollerud}, E.~J., {et~al.} 2013,
  \aap, 558, A33

\bibitem[{{Bastien et. al. in prep.}(2019)}]{2019inpreparation}
{Bastien et. al. in prep.} 2019, -,

\bibitem[{{Chambers} {et~al.}(2009){Chambers}, {Jackson}, {Rathborne}, \&
  {Simon}}]{2009ApJS..181..360C}
{Chambers}, E.~T., {Jackson}, J.~M., {Rathborne}, J.~M., \& {Simon}, R. 2009,
  \apjs, 181, 360

\bibitem[{{Chapin} {et~al.}(2013){Chapin}, {Berry}, {Gibb}, {Jenness}, {Scott},
  {Tilanus}, {Economou}, \& {Holland}}]{2013MNRAS.430.2545C}
{Chapin}, E.~L., {Berry}, D.~S., {Gibb}, A.~G., {et~al.} 2013, \mnras, 430,
  2545

\bibitem[{{Chapman} {et~al.}(2011){Chapman}, {Goldsmith}, {Pineda}, {Clemens},
  {Li}, \& {Kr{\v c}o}}]{2011ApJ...741...21C}
{Chapman}, N.~L., {Goldsmith}, P.~F., {Pineda}, J.~L., {et~al.} 2011, \apj,
  741, 21

\bibitem[{{Chen} {et~al.}(2011){Chen}, {Liu}, {Su}, \&
  {Wang}}]{2011ApJ...743..196C}
{Chen}, H.-R., {Liu}, S.-Y., {Su}, Y.-N., \& {Wang}, M.-Y. 2011, \apj, 743, 196

\bibitem[{{Cortes} {et~al.}(2008){Cortes}, {Crutcher}, {Shepherd}, \&
  {Bronfman}}]{2008ApJ...676..464C}
{Cortes}, P.~C., {Crutcher}, R.~M., {Shepherd}, D.~S., \& {Bronfman}, L. 2008,
  \apj, 676, 464

\bibitem[{{Cox} {et~al.}(2016){Cox}, {Arzoumanian}, {Andr{\'e}}, {Rygl},
  {Prusti}, {Men'shchikov}, {Royer}, {K{\'o}sp{\'a}l}, {Palmeirim}, {Ribas},
  {K{\"o}nyves}, {Bernard}, {Schneider}, {Bontemps}, {Merin}, {Vavrek}, {Alves
  de Oliveira}, {Didelon}, {Pilbratt}, \& {Waelkens}}]{2016A&A...590A.110C}
{Cox}, N.~L.~J., {Arzoumanian}, D., {Andr{\'e}}, P., {et~al.} 2016, \aap, 590,
  A110

\bibitem[{{Crutcher}(2004)}]{2004Ap&SS.292..225C}
{Crutcher}, R.~M. 2004, \apss, 292, 225

\bibitem[{{Crutcher} {et~al.}(2004){Crutcher}, {Nutter}, {Ward-Thompson}, \&
  {Kirk}}]{2004ApJ...600..279C}
{Crutcher}, R.~M., {Nutter}, D.~J., {Ward-Thompson}, D., \& {Kirk}, J.~M. 2004,
  \apj, 600, 279

\bibitem[{{Currie} {et~al.}(2014){Currie}, {Berry}, {Jenness}, {Gibb}, {Bell},
  \& {Draper}}]{2014ASPC..485..391C}
{Currie}, M.~J., {Berry}, D.~S., {Jenness}, T., {et~al.} 2014, in Astronomical
  Society of the Pacific Conference Series, Vol. 485, Astronomical Data
  Analysis Software and Systems XXIII, ed. N.~{Manset} \& P.~{Forshay}, 391

\bibitem[{{Dall'Olio} {et~al.}(2019){Dall'Olio}, {Vlemmings}, {Persson},
  {Alves}, {Beuther}, {Girart}, {Surcis}, {Torrelles}, \& {Van
  Langevelde}}]{2019A&A...626A..36D}
{Dall'Olio}, D., {Vlemmings}, W.~H.~T., {Persson}, M.~V., {et~al.} 2019, \aap,
  626, A36

\bibitem[{{Dempsey} {et~al.}(2013){Dempsey}, {Friberg}, {Jenness}, {Tilanus},
  {Thomas}, {Holland}, {Bintley}, {Berry}, {Chapin}, {Chrysostomou}, {Davis},
  {Gibb}, {Parsons}, \& {Robson}}]{2013MNRAS.430.2534D}
{Dempsey}, J.~T., {Friberg}, P., {Jenness}, T., {et~al.} 2013, \mnras, 430,
  2534

\bibitem[{{Dirienzo} {et~al.}(2015){Dirienzo}, {Brogan}, {Indebetouw},
  {Chandler}, {Friesen}, \& {Devine}}]{2015AJ....150..159D}
{Dirienzo}, W.~J., {Brogan}, C., {Indebetouw}, R., {et~al.} 2015, \aj, 150, 159

\bibitem[{{Dolginov} \& {Mitrofanov}(1976)}]{1976Ap&SS..43..291D}
{Dolginov}, A.~Z., \& {Mitrofanov}, I.~G. 1976, \apss, 43, 291

\bibitem[{{Eden} {et~al.}(2019){Eden}, {Liu}, {Kim}, {Juvela}, {Liu},
  {Tatematsu}, {Francesco}, {Wang}, {Wu}, {Thompson}, {Fuller}, {Li},
  {Ristorcelli}, {Kang}, {Hirano}, {Johnstone}, {Lin}, {He}, {Koch},
  {Sanhueza}, {Qin}, {Zhang}, {Goldsmith}, {Evans}, {Yuan}, {Zhang}, {White},
  {Choi}, {Lee}, {Toth}, {Mairs}, {Yi}, {Tang}, {Soam}, {Peretto}, {Samal},
  {Fich}, {Parsons}, {Malinen}, {Bendo}, {Rivera-Ingraham}, {Liu},
  {Wouterloot}, {Li}, {Qian}, {Rawlings}, {Rawlings}, {Feng}, {Wang}, {Li},
  {Liu}, {Luo}, {Marston}, {Pattle}, {Pelkonen}, {Rigby}, {Zahorecz}, {Zhang},
  {B{\H o}gner}, {Aikawa}, {Akhter}, {Alina}, {Bell}, {Bernard}, {Blain},
  {Bronfman}, {Byun}, {Chapman}, {Chen}, {Chen}, {Chen}, {Chen}, {Chen},
  {Chrysostomou}, {Chu}, {Chung}, {Cornu}, {Cosentino}, {Cunningham}, {Demyk},
  {Drabek-Maunder}, {Doi}, {Eswaraiah}, {Falgarone}, {Feh{\'e}r}, {Fraser},
  {Friberg}, {Garay}, {Ge}, {Gear}, {Greaves}, {Guan}, {Harvey-Smith},
  {Hasegawa}, {He}, {Henkel}, {Hirota}, {Holland}, {Hughes}, {Jarken}, {Ji},
  {Jimenez-Serra}, {Kang}, {Kawabata}, {Kim}, {Kim}, {Kim}, {Kim}, {Koo},
  {Kwon}, {Kuan}, {Lacaille}, {Lai}, {Lee}, {Lee}, {Lee}, {Li}, {Lo}, {Lopez},
  {Lu}, {Lyo}, {Mardones}, {McGehee}, {Meng}, {Montier}, {Montillaud}, {Moore},
  {Morata}, {Moriarty-Schieven}, {Ohashi}, {Pak}, {Park}, {Paladini}, {Pech},
  {Qiu}, {Ren}, {Richer}, {Sakai}, {Shang}, {Shinnaga}, {Stamatellos}, {Tang},
  {Traficante}, {Vastel}, {Viti}, {Walsh}, {Wang}, {Wang}, {Ward-Thompson},
  {Whitworth}, {Wilson}, {Xu}, {Yang}, {Yuan}, {Yuan}, {Zavagno}, {Zhang},
  {Zhang}, {Zhang}, {Zhou}, {Zhou}, {Zhu}, \& {Zuo}}]{2019MNRAS.485.2895E}
{Eden}, D.~J., {Liu}, T., {Kim}, K.-T., {et~al.} 2019, \mnras, 485, 2895

\bibitem[{{Federrath} {et~al.}(2016){Federrath}, {Rathborne}, {Longmore},
  {Kruijssen}, {Bally}, {Contreras}, {Crocker}, {Garay}, {Jackson}, {Testi}, \&
  {Walsh}}]{2016ApJ...832..143F}
{Federrath}, C., {Rathborne}, J.~M., {Longmore}, S.~N., {et~al.} 2016, \apj,
  832, 143

\bibitem[{{Foster} {et~al.}(2014){Foster}, {Arce}, {Kassis}, {Sanhueza},
  {Jackson}, {Finn}, {Offner}, {Sakai}, {Sakai}, {Yamamoto}, {Guzm{\'a}n}, \&
  {Rathborne}}]{2014ApJ...791..108F}
{Foster}, J.~B., {Arce}, H.~G., {Kassis}, M., {et~al.} 2014, \apj, 791, 108

\bibitem[{{Friberg} {et~al.}(2016){Friberg}, {Bastien}, {Berry}, {Savini},
  {Graves}, \& {Pattle}}]{2016SPIE.9914E..03F}
{Friberg}, P., {Bastien}, P., {Berry}, D., {et~al.} 2016, in \procspie, Vol.
  9914, Millimeter, Submillimeter, and Far-Infrared Detectors and
  Instrumentation for Astronomy VIII, 991403

\bibitem[{{G{\'o}mez} {et~al.}(2018){G{\'o}mez}, {V{\'a}zquez-Semadeni}, \&
  {Zamora-Avil{\'e}s}}]{2018MNRAS.480.2939G}
{G{\'o}mez}, G.~C., {V{\'a}zquez-Semadeni}, E., \& {Zamora-Avil{\'e}s}, M.
  2018, \mnras, 480, 2939

\bibitem[{{Hacar} {et~al.}(2016){Hacar}, {Alves}, {Burkert}, \&
  {Goldsmith}}]{2016A&A...591A.104H}
{Hacar}, A., {Alves}, J., {Burkert}, A., \& {Goldsmith}, P. 2016, \aap, 591,
  A104

\bibitem[{{Hacar} {et~al.}(2013){Hacar}, {Tafalla}, {Kauffmann}, \&
  {Kov{\'a}cs}}]{2013A&A...554A..55H}
{Hacar}, A., {Tafalla}, M., {Kauffmann}, J., \& {Kov{\'a}cs}, A. 2013, \aap,
  554, A55

\bibitem[{{Heitsch}(2013)}]{2013ApJ...769..115H}
{Heitsch}, F. 2013, \apj, 769, 115

\bibitem[{{Heitsch} {et~al.}(2001){Heitsch}, {Zweibel}, {Mac Low}, {Li}, \&
  {Norman}}]{2001ApJ...561..800H}
{Heitsch}, F., {Zweibel}, E.~G., {Mac Low}, M.-M., {Li}, P., \& {Norman}, M.~L.
  2001, \apj, 561, 800

\bibitem[{{Hildebrand} {et~al.}(2009){Hildebrand}, {Kirby}, {Dotson}, {Houde},
  \& {Vaillancourt}}]{2009ApJ...696..567H}
{Hildebrand}, R.~H., {Kirby}, L., {Dotson}, J.~L., {Houde}, M., \&
  {Vaillancourt}, J.~E. 2009, \apj, 696, 567

\bibitem[{{Holland} {et~al.}(2013){Holland}, {Bintley}, {Chapin},
  {Chrysostomou}, {Davis}, {Dempsey}, {Duncan}, {Fich}, {Friberg}, {Halpern},
  {Irwin}, {Jenness}, {Kelly}, {MacIntosh}, {Robson}, {Scott}, {Ade},
  {Atad-Ettedgui}, {Berry}, {Craig}, {Gao}, {Gibb}, {Hilton}, {Hollister},
  {Kycia}, {Lunney}, {McGregor}, {Montgomery}, {Parkes}, {Tilanus}, {Ullom},
  {Walther}, {Walton}, {Woodcraft}, {Amiri}, {Atkinson}, {Burger}, {Chuter},
  {Coulson}, {Doriese}, {Dunare}, {Economou}, {Niemack}, {Parsons},
  {Reintsema}, {Sibthorpe}, {Smail}, {Sudiwala}, \&
  {Thomas}}]{2013MNRAS.430.2513H}
{Holland}, W.~S., {Bintley}, D., {Chapin}, E.~L., {et~al.} 2013, \mnras, 430,
  2513

\bibitem[{{Houde} {et~al.}(2011){Houde}, {Rao}, {Vaillancourt}, \&
  {Hildebrand}}]{2011ApJ...733..109H}
{Houde}, M., {Rao}, R., {Vaillancourt}, J.~E., \& {Hildebrand}, R.~H. 2011,
  \apj, 733, 109

\bibitem[{{Houde} {et~al.}(2009){Houde}, {Vaillancourt}, {Hildebrand},
  {Chitsazzadeh}, \& {Kirby}}]{2009ApJ...706.1504H}
{Houde}, M., {Vaillancourt}, J.~E., {Hildebrand}, R.~H., {Chitsazzadeh}, S., \&
  {Kirby}, L. 2009, \apj, 706, 1504

\bibitem[{{Hull} {et~al.}(2014){Hull}, {Plambeck}, {Kwon}, {Bower},
  {Carpenter}, {Crutcher}, {Fiege}, {Franzmann}, {Hakobian}, {Heiles}, {Houde},
  {Hughes}, {Lamb}, {Looney}, {Marrone}, {Matthews}, {Pillai}, {Pound},
  {Rahman}, {Sandell}, {Stephens}, {Tobin}, {Vaillancourt}, {Volgenau}, \&
  {Wright}}]{2014ApJS..213...13H}
{Hull}, C.~L.~H., {Plambeck}, R.~L., {Kwon}, W., {et~al.} 2014, \apjs, 213, 13

\bibitem[{{Inutsuka} \& {Miyama}(1992)}]{1992ApJ...388..392I}
{Inutsuka}, S.-I., \& {Miyama}, S.~M. 1992, \apj, 388, 392

\bibitem[{{Juvela} {et~al.}(2018){Juvela}, {Guillet}, {Liu}, {Ristorcelli},
  {Pelkonen}, {Alina}, {Bronfman}, {Eden}, {Kim}, {Koch}, {Kwon}, {Lee},
  {Malinen}, {Micelotta}, {Montillaud}, {Rawlings}, {Sanhueza}, {Soam},
  {Traficante}, {Ysard}, \& {Zhang}}]{2018A&A...620A..26J}
{Juvela}, M., {Guillet}, V., {Liu}, T., {et~al.} 2018, \aap, 620, A26

\bibitem[{{Kauffmann} {et~al.}(2008){Kauffmann}, {Bertoldi}, {Bourke}, {Evans},
  \& {Lee}}]{2008A&A...487..993K}
{Kauffmann}, J., {Bertoldi}, F., {Bourke}, T.~L., {Evans}, N.~J., I., \& {Lee},
  C.~W. 2008, \aap, 487, 993

\bibitem[{{Klassen} {et~al.}(2017){Klassen}, {Pudritz}, \&
  {Kirk}}]{2017MNRAS.465.2254K}
{Klassen}, M., {Pudritz}, R.~E., \& {Kirk}, H. 2017, \mnras, 465, 2254

\bibitem[{{Koch} {et~al.}(2014){Koch}, {Tang}, {Ho}, {Zhang}, {Girart}, {Chen},
  {Frau}, {Li}, {Li}, {Liu}, {Padovani}, {Qiu}, {Yen}, {Chen}, {Ching}, {Lai},
  \& {Rao}}]{2014ApJ...797...99K}
{Koch}, P.~M., {Tang}, Y.-W., {Ho}, P.~T.~P., {et~al.} 2014, \apj, 797, 99

\bibitem[{{Lazarian} {et~al.}(1997){Lazarian}, {Goodman}, \&
  {Myers}}]{1997ApJ...490..273L}
{Lazarian}, A., {Goodman}, A.~A., \& {Myers}, P.~C. 1997, \apj, 490, 273

\bibitem[{{Li} \& {Klein}(2019)}]{2019MNRAS.485.4509L}
{Li}, P.~S., \& {Klein}, R.~I. 2019, \mnras, 485, 4509

\bibitem[{{Li} {et~al.}(2018){Li}, {Klein}, \& {McKee}}]{2018MNRAS.473.4220L}
{Li}, P.~S., {Klein}, R.~I., \& {McKee}, C.~F. 2018, \mnras, 473, 4220

\bibitem[{{Liu} {et~al.}(2019){Liu}, {Qiu}, {Berry}, {Di Francesco}, {Bastien},
  {Koch}, {Furuya}, {Kim}, {Coud{\'e}}, {Lee}, {Soam}, {Eswaraiah}, {Li},
  {Hwang}, {Lyo}, {Pattle}, {Hasegawa}, {Kwon}, {Lai}, {Ward-Thompson},
  {Ching}, {Chen}, {Gu}, {Li}, {Li}, {Liu}, {Qian}, {Wang}, {Yuan}, {Zhang},
  {Zhang}, {Zhang}, {Zhou}, {Zhu}, {Andr{\'e}}, {Arzoumanian}, {Aso}, {Byun},
  {Chen}, {Chen}, {Chen}, {Cho}, {Choi}, {Chrysostomou}, {Chung}, {Doi},
  {Drabek-Maunder}, {Dowell}, {Eyres}, {Falle}, {Fanciullo}, {Fiege},
  {Franzmann}, {Friberg}, {Friesen}, {Fuller}, {Gledhill}, {Graves}, {Greaves},
  {Griffin}, {Han}, {Hatchell}, {Hayashi}, {Hoang}, {Holland}, {Houde},
  {Inoue}, {Inutsuka}, {Iwasaki}, {Jeong}, {Johnstone}, {Kanamori}, {Kang},
  {Kang}, {Kang}, {Kataoka}, {Kawabata}, {Kemper}, {Kim}, {Kim}, {Kim}, {Kim},
  {Kim}, {Kirk}, {Kobayashi}, {Kusune}, {Kwon}, {Lacaille}, {Lee}, {Lee},
  {Lee}, {Lee}, {Liu}, {Liu}, {van Loo}, {Mairs}, {Matsumura}, {Matthews},
  {Moriarty-Schieven}, {Nagata}, {Nakamura}, {Nakanishi}, {Ohashi}, {Onaka},
  {Parker}, {Parsons}, {Pascale}, {Peretto}, {Pon}, {Pyo}, {Rao}, {Rawlings},
  {Retter}, {Richer}, {Rigby}, {Robitaille}, {Sadavoy}, {Saito}, {Savini},
  {Scaife}, {Seta}, {Shinnaga}, {Tamura}, {Tang}, {Tomisaka}, {Tsukamoto},
  {Wang}, {Whitworth}, {Yen}, {Yoo}, \& {Zenko}}]{2019ApJ...877...43L}
{Liu}, J., {Qiu}, K., {Berry}, D., {et~al.} 2019, \apj, 877, 43

\bibitem[{{Liu} {et~al.}(2018{\natexlab{a}}){Liu}, {Li}, {Juvela}, {Kim},
  {Evans}, {Di Francesco}, {Liu}, {Yuan}, {Tatematsu}, {Zhang},
  {Ward-Thompson}, {Fuller}, {Goldsmith}, {Koch}, {Sanhueza}, {Ristorcelli},
  {Kang}, {Chen}, {Hirano}, {Wu}, {Sokolov}, {Lee}, {White}, {Wang}, {Eden},
  {Li}, {Thompson}, {Pattle}, {Soam}, {Nasedkin}, {Kim}, {Kim}, {Lai}, {Park},
  {Qiu}, {Zhang}, {Alina}, {Eswaraiah}, {Falgarone}, {Fich}, {Greaves}, {Gu},
  {Kwon}, {Li}, {Malinen}, {Montier}, {Parsons}, {Qin}, {Rawlings}, {Ren},
  {Tang}, {Tang}, {Toth}, {Wang}, {Wouterloot}, {Yi}, \&
  {Zhang}}]{2018ApJ...859..151L}
{Liu}, T., {Li}, P.~S., {Juvela}, M., {et~al.} 2018{\natexlab{a}}, \apj, 859,
  151

\bibitem[{{Liu} {et~al.}(2018{\natexlab{b}}){Liu}, {Kim}, {Liu}, {Juvela},
  {Zhang}, {Wu}, {Li}, {Parsons}, {Soam}, {Goldsmith}, {Su}, {Tatematsu},
  {Qin}, {Garay}, {Hirota}, {Wouterloot}, {Chen}, {Evans}, {Graves}, {Kang},
  {Li}, {Mardones}, {Rawlings}, {Ren}, \& {Wang}}]{2018ApJ...869L...5L}
{Liu}, T., {Kim}, K.-T., {Liu}, S.-Y., {et~al.} 2018{\natexlab{b}}, \apjl, 869,
  L5

\bibitem[{{Liu} {et~al.}(2018{\natexlab{c}}){Liu}, {Kim}, {Juvela}, {Wang},
  {Tatematsu}, {Di Francesco}, {Liu}, {Wu}, {Thompson}, {Fuller}, {Eden}, {Li},
  {Ristorcelli}, {Kang}, {Lin}, {Johnstone}, {He}, {Koch}, {Sanhueza}, {Qin},
  {Zhang}, {Hirano}, {Goldsmith}, {Evans}, {White}, {Choi}, {Lee}, {Toth},
  {Mairs}, {Yi}, {Tang}, {Soam}, {Peretto}, {Samal}, {Fich}, {Parsons}, {Yuan},
  {Zhang}, {Malinen}, {Bendo}, {Rivera-Ingraham}, {Liu}, {Wouterloot}, {Li},
  {Qian}, {Rawlings}, {Rawlings}, {Feng}, {Aikawa}, {Akhter}, {Alina}, {Bell},
  {Bernard}, {Blain}, {B{\H o}gner}, {Bronfman}, {Byun}, {Chapman}, {Chen},
  {Chen}, {Chen}, {Chen}, {Chen}, {Chrysostomou}, {Cosentino}, {Cunningham},
  {Demyk}, {Drabek-Maunder}, {Doi}, {Eswaraiah}, {Falgarone}, {Feh{\'e}r},
  {Fraser}, {Friberg}, {Garay}, {Ge}, {Gear}, {Greaves}, {Guan},
  {Harvey-Smith}, {HASEGAWA}, {Hatchell}, {He}, {Henkel}, {Hirota}, {Holland},
  {Hughes}, {Jarken}, {Ji}, {Jimenez-Serra}, {Kang}, {Kawabata}, {Kim}, {Kim},
  {Kim}, {Kim}, {Koo}, {Kwon}, {Kuan}, {Lacaille}, {Lai}, {Lee}, {Lee}, {Lee},
  {Li}, {Li}, {Lo}, {Lopez}, {Lu}, {Lyo}, {Mardones}, {Marston}, {McGehee},
  {Meng}, {Montier}, {Montillaud}, {Moore}, {Morata}, {Moriarty-Schieven},
  {Ohashi}, {Pak}, {Park}, {Paladini}, {Pattle}, {Pech}, {Pelkonen}, {Qiu},
  {Ren}, {Richer}, {Saito}, {Sakai}, {Shang}, {Shinnaga}, {Stamatellos},
  {Tang}, {Traficante}, {Vastel}, {Viti}, {Walsh}, {Wang}, {Wang}, {Wang},
  {Ward-Thompson}, {Whitworth}, {Xu}, {Yang}, {Yang}, {Yuan}, {Zavagno},
  {Zhang}, {Zhang}, {Zhou}, {Zhou}, {Zhu}, {Zuo}, \&
  {Zhang}}]{2018ApJS..234...28L}
{Liu}, T., {Kim}, K.-T., {Juvela}, M., {et~al.} 2018{\natexlab{c}}, \apjs, 234,
  28

\bibitem[{{Molinari} {et~al.}(1998){Molinari}, {Brand}, {Cesaroni}, {Palla}, \&
  {Palumbo}}]{1998A&A...336..339M}
{Molinari}, S., {Brand}, J., {Cesaroni}, R., {Palla}, F., \& {Palumbo},
  G.~G.~C. 1998, \aap, 336, 339

\bibitem[{{Nakamura} \& {Li}(2008)}]{2008ApJ...687..354N}
{Nakamura}, F., \& {Li}, Z.-Y. 2008, \apj, 687, 354

\bibitem[{{Nguyen Luong} {et~al.}(2011){Nguyen Luong}, {Motte}, {Hennemann},
  {Hill}, {Rygl}, {Schneider}, {Bontemps}, {Men'shchikov}, {Andr{\'e}},
  {Peretto}, {Anderson}, {Arzoumanian}, {Deharveng}, {Didelon}, {di Francesco},
  {Griffin}, {Kirk}, {K{\"o}nyves}, {Martin}, {Maury}, {Minier}, {Molinari},
  {Pestalozzi}, {Pezzuto}, {Reid}, {Roussel}, {Sauvage}, {Schuller}, {Testi},
  {Ward-Thompson}, {White}, \& {Zavagno}}]{2011A&A...535A..76N}
{Nguyen Luong}, Q., {Motte}, F., {Hennemann}, M., {et~al.} 2011, \aap, 535, A76

\bibitem[{{Ossenkopf} \& {Henning}(1994)}]{1994A&A...291..943O}
{Ossenkopf}, V., \& {Henning}, T. 1994, \aap, 291, 943

\bibitem[{{Ostriker} {et~al.}(2001){Ostriker}, {Stone}, \&
  {Gammie}}]{2001ApJ...546..980O}
{Ostriker}, E.~C., {Stone}, J.~M., \& {Gammie}, C.~F. 2001, \apj, 546, 980

\bibitem[{{Ostriker}(1964)}]{1964ApJ...140.1056O}
{Ostriker}, J. 1964, \apj, 140, 1056

\bibitem[{{Padoan} {et~al.}(2001){Padoan}, {Goodman}, {Draine}, {Juvela},
  {Nordlund}, \& {R{\"o}gnvaldsson}}]{2001ApJ...559.1005P}
{Padoan}, P., {Goodman}, A., {Draine}, B.~T., {et~al.} 2001, \apj, 559, 1005

\bibitem[{{Pillai} {et~al.}(2015){Pillai}, {Kauffmann}, {Tan}, {Goldsmith},
  {Carey}, \& {Menten}}]{2015ApJ...799...74P}
{Pillai}, T., {Kauffmann}, J., {Tan}, J.~C., {et~al.} 2015, \apj, 799, 74

\bibitem[{{Planck Collaboration} {et~al.}(2016){Planck Collaboration}, {Ade},
  {Aghanim}, {Alves}, {Arnaud}, \& {Arzoumanian}}]{2016A&A...586A.138P}
{Planck Collaboration}, {Ade}, P.~A.~R., {Aghanim}, N., {et~al.} 2016, \aap,
  586, A138

\bibitem[{{Planck Collaboration} {et~al.}(2015){Planck Collaboration}, {Ade},
  {Aghanim}, {Alina}, {Alves}, {Armitage-Caplan}, {Arnaud}, {Arzoumanian},
  {Ashdown}, {Atrio-Barandela}, \& et~al.}]{2015A&A...576A.104P}
---. 2015, \aap, 576, A104

\bibitem[{{Poidevin} {et~al.}(2010){Poidevin}, {Bastien}, \&
  {Matthews}}]{2010ApJ...716..893P}
{Poidevin}, F., {Bastien}, P., \& {Matthews}, B.~C. 2010, \apj, 716, 893

\bibitem[{{Rathborne} {et~al.}(2011){Rathborne}, {Garay}, {Jackson},
  {Longmore}, {Zhang}, \& {Simon}}]{2011ApJ...741..120R}
{Rathborne}, J.~M., {Garay}, G., {Jackson}, J.~M., {et~al.} 2011, \apj, 741,
  120

\bibitem[{{Rathborne} {et~al.}(2006){Rathborne}, {Jackson}, \&
  {Simon}}]{2006ApJ...641..389R}
{Rathborne}, J.~M., {Jackson}, J.~M., \& {Simon}, R. 2006, \apj, 641, 389

\bibitem[{{Rathborne} {et~al.}(2008){Rathborne}, {Jackson}, {Zhang}, \&
  {Simon}}]{2008ApJ...689.1141R}
{Rathborne}, J.~M., {Jackson}, J.~M., {Zhang}, Q., \& {Simon}, R. 2008, \apj,
  689, 1141

\bibitem[{{Sakai} {et~al.}(2015){Sakai}, {Sakai}, {Furuya}, {Aikawa}, {Hirota},
  {Foster}, {Sanhueza}, {Jackson}, \& {Yamamoto}}]{2015ApJ...803...70S}
{Sakai}, T., {Sakai}, N., {Furuya}, K., {et~al.} 2015, \apj, 803, 70

\bibitem[{{Sakai} {et~al.}(2018){Sakai}, {Yanagida}, {Furuya}, {Aikawa},
  {Sanhueza}, {Sakai}, {Hirota}, {Jackson}, \&
  {Yamamoto}}]{2018ApJ...857...35S}
{Sakai}, T., {Yanagida}, T., {Furuya}, K., {et~al.} 2018, \apj, 857, 35

\bibitem[{{Sanhueza} {et~al.}(2010){Sanhueza}, {Garay}, {Bronfman}, {Mardones},
  {May}, \& {Saito}}]{2010ApJ...715...18S}
{Sanhueza}, P., {Garay}, G., {Bronfman}, L., {et~al.} 2010, \apj, 715, 18

\bibitem[{{Sanhueza} {et~al.}(2012){Sanhueza}, {Jackson}, {Foster}, {Garay},
  {Silva}, \& {Finn}}]{2012ApJ...756...60S}
{Sanhueza}, P., {Jackson}, J.~M., {Foster}, J.~B., {et~al.} 2012, \apj, 756, 60

\bibitem[{{Shepherd} {et~al.}(2004){Shepherd}, {N{\"u}rnberger}, \&
  {Bronfman}}]{2004ApJ...602..850S}
{Shepherd}, D.~S., {N{\"u}rnberger}, D.~E.~A., \& {Bronfman}, L. 2004, \apj,
  602, 850

\bibitem[{{Shepherd} {et~al.}(2007){Shepherd}, {Povich}, {Whitney},
  {Robitaille}, {N{\"u}rnberger}, {Bronfman}, {Stark}, {Indebetouw}, {Meade},
  \& {Babler}}]{2007ApJ...669..464S}
{Shepherd}, D.~S., {Povich}, M.~S., {Whitney}, B.~A., {et~al.} 2007, \apj, 669,
  464

\bibitem[{{Soam} {et~al.}(2018){Soam}, {Pattle}, {Ward-Thompson}, {Lee},
  {Sadavoy}, {Koch}, {Kim}, {Kwon}, {Kwon}, {Arzoumanian}, {Berry}, {Hoang},
  {Tamura}, {Lee}, {Liu}, {Kim}, {Johnstone}, {Nakamura}, {Lyo}, {Onaka},
  {Kim}, {Furuya}, {Hasegawa}, {Lai}, {Bastien}, {Chung}, {Kim}, {Parsons},
  {Rawlings}, {Mairs}, {Graves}, {Robitaille}, {Liu}, {Whitworth}, {Eswaraiah},
  {Rao}, {Yoo}, {Houde}, {Kang}, {Doi}, {Choi}, {Kang}, {Coud{\'e}}, {Li},
  {Matsumura}, {Matthews}, {Pon}, {Di Francesco}, {Hayashi}, {Kawabata},
  {Inutsuka}, {Qiu}, {Franzmann}, {Friberg}, {Greaves}, {Kirk}, {Li},
  {Shinnaga}, {van Loo}, {Aso}, {Byun}, {Chen}, {Chen}, {Chen}, {Ching}, {Cho},
  {Chrysostomou}, {Drabek-Maunder}, {Eyres}, {Fiege}, {Friesen}, {Fuller},
  {Gledhill}, {Griffin}, {Gu}, {Hatchell}, {Holland}, {Inoue}, {Iwasaki},
  {Jeong}, {Kang}, {Kemper}, {Kim}, {Kim}, {Lacaille}, {Lee}, {Li}, {Liu},
  {Liu}, {Moriarty-Schieven}, {Nakanishi}, {Ohashi}, {Peretto}, {Pyo}, {Qian},
  {Retter}, {Richer}, {Rigby}, {Savini}, {Scaife}, {Tang}, {Tomisaka}, {Wang},
  {Wang}, {Yen}, {Yuan}, {Zhang}, {Zhang}, {Zhou}, {Zhu}, {Andr{\'e}},
  {Dowell}, {Falle}, {Tsukamoto}, {Kanamori}, {Kataoka}, {Kobayashi}, {Nagata},
  {Saito}, {Seta}, {Hwang}, {Han}, {Lee}, \& {Zenko}}]{2018ApJ...861...65S}
{Soam}, A., {Pattle}, K., {Ward-Thompson}, D., {et~al.} 2018, \apj, 861, 65

\bibitem[{{Soler} {et~al.}(2013){Soler}, {Hennebelle}, {Martin},
  {Miville-Desch{\^e}nes}, {Netterfield}, \& {Fissel}}]{2013ApJ...774..128S}
{Soler}, J.~D., {Hennebelle}, P., {Martin}, P.~G., {et~al.} 2013, \apj, 774,
  128

\bibitem[{{Tang} {et~al.}(2019){Tang}, {Koch}, {Peretto}, {Novak},
  {Duarte-Cabral}, {Chapman}, {Hsieh}, \& {Yen}}]{2019ApJ...878...10T}
{Tang}, Y.-W., {Koch}, P.~M., {Peretto}, N., {et~al.} 2019, \apj, 878, 10

\bibitem[{{Toal{\'a}} {et~al.}(2012){Toal{\'a}}, {V{\'a}zquez-Semadeni}, \&
  {G{\'o}mez}}]{2012ApJ...744..190T}
{Toal{\'a}}, J.~A., {V{\'a}zquez-Semadeni}, E., \& {G{\'o}mez}, G.~C. 2012,
  \apj, 744, 190

\bibitem[{{Van Loo} {et~al.}(2014){Van Loo}, {Keto}, \&
  {Zhang}}]{2014ApJ...789...37V}
{Van Loo}, S., {Keto}, E., \& {Zhang}, Q. 2014, \apj, 789, 37

\bibitem[{{Wang} {et~al.}(2019){Wang}, {Lai}, {Eswaraiah}, {Pattle}, {Di
  Francesco}, {Johnstone}, {Koch}, {Liu}, {Tamura}, {Furuya}, {Onaka},
  {Ward-Thompson}, {Soam}, {Kim}, {Lee}, {Lee}, {Mairs}, {Arzoumanian}, {Kim},
  {Hoang}, {Hwang}, {Liu}, {Berry}, {Bastien}, {Hasegawa}, {Kwon}, {Qiu},
  {Andr{\'e}}, {Aso}, {Byun}, {Chen}, {Chen}, {Chen}, {Ching}, {Cho}, {Choi},
  {Chrysostomou}, {Chung}, {Coud{\'e}}, {Doi}, {Dowell}, {Drabek-Maunder},
  {Duan}, {Eyres}, {Falle}, {Fanciullo}, {Fiege}, {Franzmann}, {Friberg},
  {Friesen}, {Fuller}, {Gledhill}, {Graves}, {Greaves}, {Griffin}, {Gu}, {Han},
  {Hatchell}, {Hayashi}, {Holland}, {Houde}, {Inoue}, {Inutsuka}, {Iwasaki},
  {Jeong}, {Kanamori}, {Kang}, {Kang}, {Kang}, {Kataoka}, {Kawabata}, {Kemper},
  {Kim}, {Kim}, {Kim}, {Kim}, {Kirk}, {Kobayashi}, {Konyves}, {Kwon},
  {Lacaille}, {Lee}, {Lee}, {Lee}, {Lee}, {Li}, {Li}, {Li}, {Liu}, {Liu},
  {Lyo}, {Matsumura}, {Matthews}, {Moriarty-Schieven}, {Nagata}, {Nakamura},
  {Nakanishi}, {Ohashi}, {Park}, {Parsons}, {Pascale}, {Peretto}, {Pon}, {Pyo},
  {Qian}, {Rao}, {Rawlings}, {Retter}, {Richer}, {Rigby}, {Robitaille},
  {Sadavoy}, {Saito}, {Savini}, {Scaife}, {Seta}, {Shinnaga}, {Tang},
  {Tomisaka}, {Tsukamoto}, {van Loo}, {Wang}, {Whitworth}, {Yen}, {Yoo},
  {Yuan}, {Yun}, {Zenko}, {Zhang}, {Zhang}, {Zhang}, {Zhou}, \&
  {Zhu}}]{2019ApJ...876...42W}
{Wang}, J.-W., {Lai}, S.-P., {Eswaraiah}, C., {et~al.} 2019, \apj, 876, 42

\bibitem[{{Wang} {et~al.}(2016){Wang}, {Testi}, {Burkert}, {Walmsley},
  {Beuther}, \& {Henning}}]{wang16}
{Wang}, K., {Testi}, L., {Burkert}, A., {et~al.} 2016, \apjs, 226, 9

\bibitem[{{Ward-Thompson} {et~al.}(2017){Ward-Thompson}, {Pattle}, {Bastien},
  {Furuya}, {Kwon}, {Lai}, {Qiu}, {Berry}, {Choi}, {Coud{\'e}}, {Di Francesco},
  {Hoang}, {Franzmann}, {Friberg}, {Graves}, {Greaves}, {Houde}, {Johnstone},
  {Kirk}, {Koch}, {Kwon}, {Lee}, {Li}, {Matthews}, {Mottram}, {Parsons}, {Pon},
  {Rao}, {Rawlings}, {Shinnaga}, {Sadavoy}, {van Loo}, {Aso}, {Byun},
  {Eswaraiah}, {Chen}, {Chen}, {Chen}, {Ching}, {Cho}, {Chrysostomou}, {Chung},
  {Doi}, {Drabek-Maunder}, {Eyres}, {Fiege}, {Friesen}, {Fuller}, {Gledhill},
  {Griffin}, {Gu}, {Hasegawa}, {Hatchell}, {Hayashi}, {Holland}, {Inoue},
  {Inutsuka}, {Iwasaki}, {Jeong}, {Kang}, {Kang}, {Kang}, {Kawabata}, {Kemper},
  {Kim}, {Kim}, {Kim}, {Kim}, {Kim}, {Kim}, {Lacaille}, {Lee}, {Lee}, {Li},
  {Li}, {Liu}, {Liu}, {Liu}, {Liu}, {Lyo}, {Mairs}, {Matsumura},
  {Moriarty-Schieven}, {Nakamura}, {Nakanishi}, {Ohashi}, {Onaka}, {Peretto},
  {Pyo}, {Qian}, {Retter}, {Richer}, {Rigby}, {Robitaille}, {Savini}, {Scaife},
  {Soam}, {Tamura}, {Tang}, {Tomisaka}, {Wang}, {Wang}, {Whitworth}, {Yen},
  {Yoo}, {Yuan}, {Zhang}, {Zhang}, {Zhou}, {Zhu}, {Andr{\'e}}, {Dowell},
  {Falle}, \& {Tsukamoto}}]{2017ApJ...842...66W}
{Ward-Thompson}, D., {Pattle}, K., {Bastien}, P., {et~al.} 2017, \apj, 842, 66

\bibitem[{{Wardle} \& {Kronberg}(1974)}]{1974ApJ...194..249W}
{Wardle}, J.~F.~C., \& {Kronberg}, P.~P. 1974, \apj, 194, 249

\bibitem[{{Wareing} {et~al.}(2016){Wareing}, {Pittard}, {Falle}, \& {Van
  Loo}}]{2016MNRAS.459.1803W}
{Wareing}, C.~J., {Pittard}, J.~M., {Falle}, S.~A.~E.~G., \& {Van Loo}, S.
  2016, \mnras, 459, 1803

\bibitem[{{Xu} {et~al.}(2016){Xu}, {Li}, {Zhang}, {Liu}, {Wang}, {Ning}, \&
  {Ju}}]{2016ApJ...819..117X}
{Xu}, J.-L., {Li}, D., {Zhang}, C.-P., {et~al.} 2016, \apj, 819, 117

\bibitem[{{Yanagida} {et~al.}(2014){Yanagida}, {Sakai}, {Hirota}, {Sakai},
  {Foster}, {Sanhueza}, {Jackson}, {Furuya}, {Aikawa}, \&
  {Yamamoto}}]{2014ApJ...794L..10Y}
{Yanagida}, T., {Sakai}, T., {Hirota}, T., {et~al.} 2014, \apjl, 794, L10

\bibitem[{{Zhang} {et~al.}(2014){Zhang}, {Qiu}, {Girart}, {Liu}, {Tang},
  {Koch}, {Li}, {Keto}, {Ho}, {Rao}, {Lai}, {Ching}, {Frau}, {Chen}, {Li},
  {Padovani}, {Bontemps}, {Csengeri}, \& {Ju{\'a}rez}}]{2014ApJ...792..116Z}
{Zhang}, Q., {Qiu}, K., {Girart}, J.~M., {et~al.} 2014, \apj, 792, 116

\end{thebibliography}

\end{document}